\documentclass[12pt,preprint]{aastex}

\def\spose#1{\hbox to 0pt{#1\hss}}
\def\approxlt{\mathrel{\spose{\lower 3pt\hbox{$\sim$}}
        \raise 2.0pt\hbox{$<$}}}
\def\approxgt{\mathrel{\spose{\lower 3pt\hbox{$\sim$}}
        \raise 2.0pt\hbox{$>$}}}

\def\multleft#1{\hbox to size{\vbox {\halign {\lft{##}\cr #1}}\hfill}\par}
\def\multright#1{\hbox to size{\vbox {\halign {\rt{##}\cr #1}}\hfill}\par}

\def\boxit#1{\vbox{\hrule\hbox{\vrule\kern3pt\vbox{\kern3pt
          #1 \kern3pt}\kern3pt\vrule}\hrule}}

\def\cm{{\rm\thinspace cm}}

\def\erg{{\rm\thinspace erg}}

\def\km{{\rm\thinspace km}}

\def\Msun{\hbox{$\rm\thinspace M_{\odot}$}}

\def\s{{\rm\thinspace s}}

\def\chisq{\hbox{$\chi^2$}}

\def\pcmsq{\hbox{$\cm^{-2},$}}

\def\ergpcmsqps{\hbox{$\erg\cm^{-2}\s^{-1}\,$}}

\def\ergps{\hbox{$\erg\s^{-1}\,$}}

\def\kmps{\hbox{$\km\s^{-1}\,$}}

\def\pcm{\hbox{$\cm^{-3}\,$}}
\def\pcmsq{\hbox{$\cm^{-2}\,$}}

\slugcomment{submitted to ApJ} 
\shorttitle{Spatially resolved X-ray spectra of NGC 4258}
\shortauthors{Yang et al.}

\begin{document}
\title{Spatially resolved X-ray spectra of NGC~4258}
\author{Y. Yang, B. Li, A. S. Wilson, C. S. Reynolds}
\affil{Department of Astronomy, University of Maryland, College Park 20742}
\email{yyang@astro.umd.edu}
\begin{abstract}
We report a spatially resolved, X-ray spectral analysis of NGC~4258 using archival {\it Chandra} and {\it XMM-Newton} observations. The {\it XMM-Newton} spectra of the nuclear region are well described by two power-law components, a soft (0.57 keV) thermal component, and an Fe K$\alpha$ line with EW = 40 $\pm$ 33~eV.  The properties of the second, weaker power-law component are similar to those of an off-nuclear source $2.5\arcsec$ SW of the nucleus.  The spectrum of the extended emission of the entire galaxy is well described by two thermal components (MEKAL) models with temperatures $\simeq 0.60$ and 0.22~keV. The {\it Chandra} and {\it XMM-Newton} spectra along the anomalous arms show that the absorbing column density to the SE anomalous arm is consistent with absorption by gas in our Galaxy, while the absorbing column to the NW anomalous arm is higher, indicating that the NW arm is partially on the far side of the galactic disk.  The combined {\it Chandra} data clearly detect the X-ray emission from the  hot spots at the end of the approximately N-S radio jets.  By assuming  the hot spots represent shocked thermal gas at the ends of the jets, we estimate shock powers of $\simeq 3 \times 10^{39} f^{-1/2}$~\ergps ($f $ is the filling factor), similar to the radiative power in the inner anomalous arms, consistent with the notion that the jets could be responsible for heating the gas in the anomalous arms.    
\end{abstract}


\section{Introduction}
The interaction between nuclear activity and host galaxy is a very interesting problem in the study of active galactic nuclei (AGNs).  Such interactions could be rather strong as in the case of NGC~4258 (M~106),  a nearby (7.2 Mpc, \markcite{herrnstein99}{Herrnstein} {et~al.} 1999) prominent barred Seyfert 1.9  galaxy.  NGC~4258 is one of the best laboratories to study accretion onto blackholes because of its well determined blackhole mass and distance. The spatial--velocity distribution of water mega-maser sources in NGC~4258 on scales of 0.14--0.28~pc indicates a thin Keplerian disk rotating around a putative blackhole with mass $3.9 \times 10^7$~M$_{\odot}$ \markcite{miyoshi95, herrnstein99}({Miyoshi} {et~al.} 1995; {Herrnstein} {et~al.} 1999). The accretion disk has an almost edge on orientation with the rotation axis pointing almost north (PA = -8\arcdeg) and tipped $82\arcdeg$ to the line-of-sight . A pc scale jet closely aligns in projection on the sky with the rotation axis \markcite{herrnstein97}({Herrnstein} {et~al.} 1997).  

The galaxy is well known for its pair of ``anomalous arms'',  located between the normal spiral arms of the galaxy, and visible in  H$\alpha$, soft X-ray, and radio continuum. These arms emerge from the nuclear region as linear features, then bend at 2-5~kpc radius in a trailing sense with respect to the galactic rotation. The smooth appearance and the lack of blue light from young stars make them distinctive from the normal spiral arms. Many authors suggest that the anomalous arms are ejected from the nucleus (e. g. \markcite{vanderkruit72, vanalbada77, icke79}{van der Kruit}, {Oort}, \&  {Mathewson} 1972; {van Albada} 1977; {Icke} 1979),  in the forms of either ``cannon balls" or jets.  However, there is strong evidence that at least the inner part of the anomalous arms lie in the disk of NGC 4258 ({\markcite{vanderkruit74, vanalbada82, hummel89, plante91, cecil92}{van der Kruit} 1974; {van Albada} \& {van der Hulst} 1982; {Hummel}, {Krause}, \& {Lesch} 1989; {Plante} {et~al.} 1991; {Cecil}, {Wilson}, \& {Tully} 1992). It is hard to understand how the ejecta could be confined to the thin galactic disk  while propagating to the observed distances of  $> 2 $~kpc. The energy requirement to push aside the dense interstellar gas in the disk is also unrealistically high. A more serious problem with the jet models is that  a pair of currently active jets, which are extensions of the parsec scale jet seen in VLBA images and which extend to projected distances of 870~pc south and 1.7~kpc north of the nucleus, is clearly visible in the VLA image \markcite{cecil00}({Cecil} {et~al.} 2000).  The galaxy disk has an inclination of 64\arcdeg, while the radio jet lies very close to the polar axis of the maser disk, which itself is inclined by 60$^{\circ}$ to the polar axis of the galaxy stellar disk. This misalignment between the currently active jets and the anomalous arms makes the jet models for the anomalous arms unlikely. 
X-ray emission from the anomalous arms was first discovered with the {\it Einstein} HRI \markcite{cecil92}({Cecil} {et~al.} 1992).  Subsequent observations from {\it ROSAT} \markcite{pietsch94, cecil95, vogler99}({Pietsch} {et~al.} 1994; {Cecil}, {Morse}, \& {Veilleux} 1995; {Vogler} \& {Pietsch} 1999) have found that the X-ray emission is closely associated with the anomalous arms, and is thermal in nature. The temperature of the hot gas was found to be $kT \simeq 0.3 - 0.6$~keV, with a possible halo component \markcite{vogler99}({Vogler} \& {Pietsch} 1999).  Combining the results from VLA and VLBA, and  with the help of a 14~ks {\it Chandra} observation, \markcite{wilson01}{Wilson}, {Yang}, \& {Cecil} (2001, hereafter WYC) showed that the vertical projection of the jets onto the galactic disk aligns closely with the inner anomalous arms, suggesting the anomalous arms may be  shock heated gas powered by hot halo gas driven into the galactic disk by the out-of-plane jets. To better understand the nature of the anomalous arms, their relation with the jets and the AGN, we need detailed knowledge of the X-ray emitting gas in the anomalous arms, the jets and the nucleus. Such knowledge will allow constraints on the density, pressure, and energetics in the jets and the anomalous arms. 

The nucleus of NGC~4258 is heavily obscured in the {\it ROSAT} band, and the first hard X-ray detection was reported using {\it ASCA} observations, with the absorption column density $N_H \simeq 1.5 \times 10^{23}$~\pcmsq.  A narrow 6.4 keV Fe K$\alpha$ line was detected with an equivalent width (EW)  of $250 \pm 100$~eV \markcite{makishima94}({Makishima} {et~al.} 1994). \markcite{fruscione05}{Fruscione} {et~al.} (2005) noticed that the column density remained mostly constant ($\sim 10^{23}$~\pcmsq) over 9 years, except  on two occasions, which they attribute to absorption by a warped accretion disk.  On the other hand, flux variability in the 2--10~keV X-ray band has been seen on timescales from 19 days to years, with little change in the power-law index \markcite{reynolds00, terashima02, fruscione05}({Reynolds}, {Nowak}, \&  {Maloney} 2000; {Terashima} {et~al.} 2002; {Fruscione} {et~al.} 2005).  The Fe K$\alpha$ line was found to be weaker in recently reported {\it ASCA} \markcite{reynolds00}({Reynolds} {et~al.} 2000), {\it BeppoSax} \markcite{fiore01}({Fiore} {et~al.} 2001) and {\it RXTE} (Mushotzky, private communication) observations,  with  EW$\simeq 100$~eV.  Recent observations from {\it XMM} tend to find much lower EW, and in many cases, only upper limits \markcite{pietsch02,  fruscione05}({Pietsch} \& {Read} 2002; {Fruscione} {et~al.} 2005). Moreover, weak Fe K$\alpha$ {\it absorption} features were reported using a {\it Chandra} observation \markcite{young04}({Young} \& {Wilson} 2004), but not detected in {\it XMM-Newton} observations.  The variability of EW has been confirmed in a recent reanalysis of the archival {\it ASCA}, {\it RXTE} and {\it XMM} data, using the latest calibrations (Hanson et al. 2006, submitted; hereafter Hanson06). 

In this paper, we present our analysis of the archival {\it Chandra} and {\it XMM-Newton} observations of NGC~4258, with a focus on the anomalous arms and jets.  With the excellent spatial resolution of {\it Chandra} and the sensitivity of {\it XMM-Newton},  we are able to probe X-ray emitting gas in the anomalous arms and jet hot spots in detail. In \S~\ref{data}, we first present the data reduction. The results are presented in  \S~\ref{results}. We discuss the implications of the results in \S~\ref{discussion},  and give concluding remarks in \S~\ref{conclusion}.  

\section{Data reduction
\label{data}}
\subsection{{\it Chandra} observations}
NGC~4258 was observed with the Advanced CCD Image Spectrometer (ACIS) on {\it Chandra}
on April 17, 2000 and May 28, 2001 (Table~\ref{tab1}). The nucleus was located at the aim point on the S3 chip.
The spatial resolution in the  0.4--2 keV band is $< 1\arcsec$ for all the off-axis angles covered
by the detected diffuse emission from NGC~4258.  The default frame time of 3.2~s was used. 
The data were reduced using CIAO~3.3 with calibration version CALDB 3.2. We followed the standard procedure in reducing the data. By comparing the positions of the nucleus and point sources which have radio counterparts in the 6~cm high resolution {\it Very Large Array} (VLA) image used in \markcite{cecil00}{Cecil} {et~al.} (2000), we found the astrometric error of the {\it Chandra} image is $< 0.5\arcsec$. 

In Fig.~\ref{true_color}, we show the adaptively smoothed ``true color" image of NGC~4258 in the 0.4 -- 2~keV band.  The anomalous arms  appears smooth but not featureless.  The northern hot spot (marked ``N") of the radio jet is clearly visible in the X-ray image. An arc-like structure, suggestive of a shock wave,  is seen $\sim 6\arcsec$ to the west of the northern hot spot (N) in the northern anomalous arm,  indicating a possible relation with the hot spot.  The arc is $\sim 1\arcmin$  from the  nucleus. A similar ``arc" is also visible, but less prominent, in the southern anomalous arm,  at about the same distance from the nucleus as the northern arc.  The X-ray color is not uniform along the anomalous arms and the surrounding regions. It is obvious that the SE arm is in general redder (softer)  than the NW arm. The diffuse emission surrounding the anomalous arms  is generally soft, with the emission from the NE quadrant redder (softer) than that in the SW quadrant. 

We extract spectra in regions along the anomalous arms as well as the hot spots which are detected in X-rays (Fig.~\ref{regions}). The regions are chosen so that there is little variation of  color within each region.  In the presence of large scale X-ray emission, extra care is needed in determining the level of background. We compare the following methods. First, we visually inspect the image and extract  the background spectrum from two regions with no point sources and which show little diffuse emission.  The source spectrum from a region on the anomalous arms with low surface brightness is also extracted. The resulting two background subtracted spectra using the two background files are compared,  and no significant difference was found. Second, we extract background spectra at the same physical location as the source from a stack of background files compiled by Maxim Markevitch\footnote{Available at http://cxc.harvard.edu/contrib/maxim/acisbg/} , which has been reprojected to the same sky coordinates as our observations.  The resulting background subtracted spectra is compatible with those using the first method.  

\subsection{{\it XMM-Newton} Observations}
We use the data from five {\it XMM-Newton} observations of NGC~4258 available in the archive (Table~\ref{tab1}). The data is reprocessed using the latest calibration files and SAS 6.5.0. The resulting event lists from the two MOS detectors and the PN detector are filtered for background flares, using visual inspection of the light curves of the whole field.  The resulting good exposure times can be found in Table~\ref{tab1}. Comparing with the filtered exposure time of \markcite{fruscione05}{Fruscione} {et~al.} (2005), which uses a signal-to-noise criterion, we found good agreements. The total good time from the observations are 132.4~ks for the two MOS detectors combined and 45.3~ks for the PN detector.   

We extract spectra along the anomalous arms and from the regions shown in Fig.~\ref{regions}, which are determined using the {\it Chandra} data (left image). The background spectra are taken from regions of low emission on the same chip close to  the ``source" regions. The source region sizes are large enough that the number of events from neighboring regions due to the point spread function (PSF) wings is small, except in regions close to the nucleus (particularly regions 10 and 11), where scattered light from the bright nucleus affects the spectra of the diffuse emission.  However, this ``mixing" does not seem to alter the qualitative conclusions for these regions (see \S~\ref{extended}). The effect can be modeled using {\it xmmpsf} in XSPEC. Unfortunately,  the PI spectra produced by the current version of SAS are not compatible with the {\it xmmpsf} model.  XSELECT included in the {\it ftools} package can produce spectra that are compatible with {\it xmmpsf}, but only spectra from a single instrument from a single observation can be treated properly with {\it xmmpsf}. Thus we rely more on comparing the {\it XMM-Newton} spectra with those from {\it Chandra}.  

The spectra from different observations which use the same detector (MOS or PN) are co-added using the {\it addspec} task in the {\it ftools} package.  The routine combines the spectra from the source, background and produces a new response matrix.  The spectra from both MOS detectors are also added. The addition of the spectra of the extended emission is justified because the emission is not variable. 

While the cross calibration between PN and MOS has improved significantly since the release of SAS 6.5.0, the calibration is still not very good below 0.8~keV and above 7 keV\footnote{see Kirsh (2006), ``EPIC status of calibration and data analysis" (CAL-TN-0018).}. This systematic uncertainty can cause spectral fitting  to produce larger $\chisq$ values if the spectra from both PN and MOS are fitted simultaneously. Experimenting with PN and MOS spectra, we found that the MOS spectra tend to produce more artifacts that mimic emission lines in the 0.2--2 keV band. In the case of thermal gas with temperature $\sim 0.5$~keV, fitting MOS spectra produces higher abundance values. In this study, we present the spectral parameters using only PN as well as using both PN and MOS detectors. 
     
\section{Results
\label{results}}
\subsection{The nuclear region}
The spectrum of the nuclear region has previously been reported using {\it XMM-Newton} (\markcite{pietsch02, fruscione05}{Pietsch} \& {Read} 2002; {Fruscione} {et~al.} 2005, Hanson06) and {\it Chandra} \markcite{young04, fruscione05}({Young} \& {Wilson} 2004; {Fruscione} {et~al.} 2005).  These observations, as well as the earlier {\it ASCA} observations \markcite{reynolds00}({Reynolds} {et~al.} 2000), have produced two puzzles: 1) A weak second power-law component is needed in addition to the heavily absorbed power-law. The nature of this component is unknown. It could come from scattered light from the nucleus, or from unresolved point sources; 2) Weak absorption feature were identified in {\it Chandra} observations but not detected in {\it XMM-Newton} observations (\markcite{fruscione05}{Fruscione} {et~al.} 2005 reject these detections, but the issue seems to lie in the method of statistics).  

As noticed in WYC and Young \& Wilson (2004), a bright off-nuclear source (J121857.3+471812) $2.5\arcsec$ to the SW of the nucleus  has been found in the {\it Chandra} observations. Because the source
is significantly weaker than the nucleus, it is commonly assumed that it contributes little to the spectra obtained with instruments which cannot separate the point source from the nucleus, and it is thus ignored.  In this section, we re-examine the spectra of the nuclear region and the off-nuclear source, and discuss the effects of the off-nuclear source on the {\it XMM-Newton} spectrum in which the off-nuclear source is not resolved from the nucleus.  We then compare the {\it Chandra} spectra of the off-nuclear source and that of the nuclear region from {\it XMM-Newton}. In contrast to previous works, we use stacked spectra to improve the signal-to-noise ratio. 

\subsubsection{The {\it XMM-Newton} Spectra of the nuclear region}
The spectrum in each observation is extracted from a circular region centered at the nucleus with a radius of $20\arcsec$.  To obtain meaningful results from a stacked spectrum, either the source or the response of the instrument has to be invariable.  While little variability is seen below 2~keV, the spectra of the nucleus show moderate variability above this energy. We check the variation of the EPIC responses using simulations. The response matrices  produced for each of the observations are convolved with the  best-fit model from the longest {\it XMM-Newton} observation.  Above 2~keV, there is no difference between the convolved spectra, indicating little change in the response in the hard band. Below 1 keV, on the other hand, there is a small change in the response. Fortunately, the spectra in this band show no variability. This validates the stacking  method.  
 
We first study the stacked PN spectrum in the 0.2--10~keV band. The spectrum is grouped to a minimum of 25 counts per bin to allow use of $\chisq$ fitting. We employ Model J from Hanson06, which is composed of two  power-law components,  a collisionally ionized thermal component based on the MEKAL model \markcite{mewe85, mewe86,liedahl95}({Mewe}, {Gronenschild}, \& {van den  Oord} 1985; {Mewe}, {Lemen}, \& {van den Oord} 1986; {Liedahl}, {Osterheld}, \&  {Goldstein} 1995), and an Fe K$\alpha$ emission line.  The first power-law component comes from the AGN and is more obscured. The second power-law is much weaker than the first one, and is assumed to be unobscured by the gas shrouding the AGN. The best-fit parameters and the 90\% confidence ranges for a single interesting parameter (in small numbers next to the parameters) are listed  in Table~\ref{tab_nucleus} (Fit 1). The aperture corrected model 2--8 keV flux is  $(7.1 \pm 0.4) \times 10^{-12}$~\ergpcmsqps ($1\sigma$ error). The corresponding unabsorbed luminosity is $4.4 \times 10^{40}$~\ergps.  

To  better constrain the Fe K$\alpha$ emission, we use both the PN and MOS spectra above 2.2~keV (Fig.~\ref{nuc_spec}). Choosing only the hard band emission allows us to use a single absorbed power-law as the baseline model of the continuum. We obtain $N_H =(8.2 \pm 0.3) \times 10^{22}$\pcmsq,  $\Gamma= 1.57 \pm 0.05$ 
\footnote{The errors represent the 90\% confidence ranges for a single interesting parameter. This convention is used throughout this paper unless noted otherwise.}. A narrow iron K$\alpha$ emission line ($E_{line} = 6.39 \pm 0.04$~keV, $ \sigma < 0.1$~keV) is clearly seen in both the PN and MOS spectra. By adding a narrow emission line (3 degrees of freedom, dof.) $\chisq$ improves from 992.5 to 969.2, corresponding to a confidence level of  99.996\%  according to an F-test.  The equivalent width of the line is $40 \pm 33$~eV.  The line flux is   $3.5^{5.2}_{2.3}  \times 10^{-6}$ photons cm$^{-2}$ s$^{-1}$. These values agree very well with those from Hanson06, in which the PN data from the five observations have been fitted simultaneously,  but with the absorption left  free for each spectrum. The properties of the Fe K$\alpha$ emission line also agrees with the upper limit found in  \markcite{fruscione05}{Fruscione} {et~al.} (2005). 

\subsubsection{Off-nuclear source
\label{s_off_nuc}}
The off-nuclear source J121857.3+471812 was first reported in WYC based on the {\it Chandra} April 17, 2000 observation. The {\it Chandra} spectrum from the May 28, 2001 observation has also been reported by \markcite{young04}{Young} \& {Wilson} (2004). We extract the {\it Chandra} spectra of the source from a circular region with radius of $\sim 1.5\arcsec$ (which corresponds to an enclosed energy of 95\% at 1.4~keV ). The spectra of the background have been extracted from a region at the same distance from the nucleus but on the opposite side of the off-nuclear source.  This allows the PSF scattered nuclear photons in the source extraction region to be properly taken into account. Spectra from both observations can be described very well with a  single absorbed power-law, and the resulting parameters agree well with the previously reported values.  There is no significant variability between the two epochs.  This allows us to fit the combined spectra to improve the signal-to-noise ratio. The resulting best-fit parameters are $N_H = 3.4_{2.4}^{5.0} \times 10^{21}$~\pcmsq, $\Gamma = 1.74_{1.52}^{2.02}$, $f_{0.5-8~keV} =  1.2_{0.8}^{1.3} \times 10^{-13}$~\ergpcmsqps. If the source is in NGC~4258, its 0.5 -- 8~keV luminosity  is $\sim 7 \times 10^{38}$~\ergps.

We compare the flux from the off-nuclear source with that of the second power-law component in the {\it XMM-Newton} spectra.  We replace the second power-law in FIT~1 (Table~\ref{tab_nucleus}) with an absorbed power-law with the same index and absorption column density as those of the off-nuclear source, found from the {\it Chandra} observations.   The resulting best-fit parameters are shown in Table~\ref{tab_nucleus} as FIT~2. The $\chisq$ increases slightly in this new fit but is still acceptable.  The new fit is still better than that without introducing the second power-law component.  The  0.5--8~keV flux of the second power-law component is $\sim 1.4_{1.0}^{1.6} \times 10^{-13}$~\ergpcmsqps, which agrees very well with the flux of the off-nuclear source from {\it Chandra}.  This suggests  that the off-nuclear source is likely to be responsible for  the second power-law component in the {\it XMM-Newton} spectrum and leaves less room for scattered light from the nucleus. 

We also search for emission lines in the combined spectra of the off-nuclear source. Given the small number of photons in these observations, we regroup the spectrum to $>3$ counts per bin.  Interestingly, an emission line like feature at $\sim 6.4$~keV is seen in this rather noisy spectrum (Fig.~\ref{fig_offnuc}).  Unfortunately, only the  May 28, 2001 observation has enough counts above 5~keV to show this feature.  By adding an unresolved Fe K$\alpha$ emission line to the continuum model, and  using the C-statistic  \markcite{cash79}({Cash} 1979),  we found the energy of the line feature to be $E_{line} = 6.36^{ 6.49}_{ 6.26}$~keV ($1\sigma$ confidence range) and an upper limit to the line flux  of $1.6 \times 10^{-6}$~photons cm$^{-2}$ s$^{-1}$.  To check if the line could be caused by X-ray photons from the nucleus scattered by the HRMA, we examine the stacked {\it Chandra} image extracted between 6.2 and 6.6 keV.  A small number of events can be seen around the nucleus. Within the spectrum extraction region of the off-nuclear source, 3 events are found, two of which are in the same pixel.  In an annular region with $r_{in} = 1\arcsec$, $r_{out} = 4\arcsec$  and which is  centered on the nucleus and  overlaps the off-nuclear source extraction circle,  we found 7 counts in the 6.2 to 6.6~keV band, all of which are in the SW quadrant with respect to the nucleus (i.e. in the direction of the off-nuclear source).  Comparing the counts in the annulus and the source extraction region,  we found the significance level of detection of  the off-nuclear source is $\sim 84\%$ between 6.2 and 6.6 keV (we have used the  best-fit upper-limits in \markcite{gehrels86}{Gehrels} (1986) to calculate the significance level).  We therefore cannot rule out the possibility that the line represents HRMA scattered nuclear light.  However, given that no strong Fe K$\alpha$ emission line is detected in the nuclear spectrum in the May 28, 2001 {\it Chandra} observation, and that the photons detected around the line energy seem to ``tilt" toward the off-nuclear source, it is very possible that the emission line does come from the off-nuclear source.  It is noticeable that the upper limit to its  line flux from the {\it Chandra} spectrum is only a factor of 2 lower than the Fe K$\alpha$ emission line flux of the entire nuclear region from {\it XMM-Newton}.  Should the line be confirmed to come from the off-nuclear source,  it would certainly complicate the interpretation of the Fe K$\alpha$ line results from instruments that cannot resolve the off-nuclear source from the nucleus. An emission line from the off-nuclear source might also ``fill-in" one of the putative absorption lines from the nucleus, such as those reported in \markcite{young04}{Young} \& {Wilson} (2004), making the absorption line harder to detect with {\it XMM-Newton}.  Unfortunately,  Fe K$\alpha$ absorption from the nucleus was only reported in only the April 17, 2000 observation \markcite{young04}({Young} \& {Wilson} 2004). This prevents a direct check on how the tentative emission line from the off-nuclear source would affect the spectrum of the nucleus if the off-nuclear source were not resolved from the nucleus, as is the case in the {\it XMM-Newton} observations.     

\subsection{The Hot gas in NGC~4258:  Anomalous arms and other diffuse components
\label{extended}}
As shown by {\it ROSAT} observations \markcite{cecil95, vogler99}({Cecil} {et~al.} 1995; {Vogler} \& {Pietsch} 1999),  the extended emission in NGC 4258 is predominantly thermal.  \markcite{vogler99}{Vogler} \& {Pietsch} (1999) found that the emission can be decomposed into two thermal components, with temperatures of $\sim 0.5$ and 0.2~keV.  The absorption corrected total luminosity $L_{0.1-2.4~keV} \sim 2 \times 10^{40}$~\ergps. The low temperature component is suggested to originate from a halo component, similar to those found in other spiral galaxies. 

We first look at the global spectrum of the galaxy. We extract the {\it  XMM-Newton} spectra from an elliptical region centered on the nucleus, with semi-major and minor axes of $2.6\arcmin$ and $1.5\arcmin$,  respectively. The major axis of the ellipse is in the direction of the anomalous arms and is close to the major axis of the galaxy disk. The extraction region covers most of the soft emission while avoiding any bright  point sources.  Because the low energy tail of the absorbed power-law emission from the nucleus blends with the thermal emission around 2 keV, we model the spectra over 0.2--10~keV.   We first test a model similar to the one we used to fit the nuclear spectra, which consists of two power-laws, a MEKAL, and an Fe K$\alpha$ emission line. The model produces a very good fit to the spectrum above 2 keV, and agrees very well with the best-fit to the nucleus.  Below 2~keV, a large residual from this model is clearly seen, indicating the emission cannot be well described with a single thermal emission component.   We add a second  MEKAL component.  Unfortunately, the abundance of this component is poorly constrained by the data, and the best-fit abundance is unrealistically high. We thus choose a fixed abundance equal to 0.15 of the solar value, a typical value found in the outer regions of the extended emission (see below).  The addition of the second MEKAL component significantly improve the fit.  The temperatures of the best-fit MEKAL models are  $0.595 \pm 0.005$ and $ 0.22 \pm 0.01$~keV.  This confirms  the  results of \markcite{vogler99}{Vogler} \& {Pietsch} (1999).  The 0.2--2.4~keV fluxes are found to be $(2.38 \pm 0.05) \times 10^{-12}$~\ergpcmsqps for the 0.59 keV component ,  and $(5.10 \pm 0.06) \times 10^{-13}$ for the 0.22~keV component. The total absorption corrected luminosity is $L_{0.1-2.4~keV} \sim 1.7 \times 10^{40}$~\ergps.   
  
The spectra in the regions shown in Fig.~\ref{regions} are investigated using both {\it Chandra} and {\it XMM-Newton} data.  We first model the extended emission with a single thermal component, which produces a very good fit  to all the {\it Chandra} spectra between 0.4 and 2~keV.   In all regions except  6, 10, and 11,  the {\it XMM-Newton} spectra in the 0.2--2 keV band are used. In these three regions,  the hard band emission from the AGN contributes significantly to the 0.2--2~keV spectra. We thus model the spectra in regions 6, 10, and 11 between 0.2 and 10.0~keV using a MEKAL  plus an absorbed power-law model.  In Table~\ref{tab_aa}, we list the best-fit parameters using {\it XMM-Newton} spectra. While producing consistent results,  including the MOS spectra seems to systematically  decrease the best-fit absorption column density and increase the temperature. The reduced $\chisq$ per dof is also higher if the MOS spectra are included.  This is still true even if we allow the normalization of the models to be different for the two instruments.  

The best-fit absorption column densities and temperatures of the thermal gas along the anomalous arms are shown in Fig. \ref{aarm_NH_KT}.   It is obvious that $N_H$ increases along  the anomalous arms from SE to NW.  The best-fit column densities in the SE arm are consistent with that of the Milky Way Galaxy in the direction of NGC~4258, indicating little absorption intrinsic to NGC~4258 in the SE arm.  The column densities in the NW arm, on the other hand,  are a factor of 3--4 times the Galactic value. The temperature, on the other hand,  peaks at the nucleus (Region 6) and decreases slowly as a function of distance from the nucleus in both NW and SE directions.  There is no systematic difference in the temperature distribution between SE and NW arms.  The increase of column density is naturally interpreted if the NW arm is slightly behind the galactic disk of NGC~4258, as argued by WYC.  The SE arm, on the other hand, is either in the galactic plane, or  in front of it.

The emissions in regions 10 and 11, which are located to the NE and SW of the nucleus,  are contaminated by scattered nuclear light from the broad PSF wing of {\it XMM-Newton}. This is much less of a problem for {\it Chandra}.  We first examine the {\it XMM-Newton} spectra of Regions 10 and 11 without using the {\it xmmpsf} model.  There is little difference in the best-fit temperature between the two regions (Table~\ref{tab_aa}). On the other hand, the absorption column density in Region 11 is about twice the value of Region 10.  Joint analysis of the spectra suggest the difference in the column density is at $>4\sigma$ level.  Given the inclusion of scattered nuclear light  in both spectra, the difference in column density is likely to be higher. Modeling  0.4--2 keV {\it Chandra} spectra of the two regions yields $N_H =1.7_{0.0}^{8.8} \times 10^{20}$~\pcmsq, $kT = 0.49_{0.38}^{0.54}$~keV for Region 10 and $N_H =7.8_{3.5}^{14.} \times 10^{20}$~\pcmsq,  $kT = 0.57_{0.49}^{0.61}$~keV for region 11, which are consistent with the results using {\it XMM-Newton} (the abundance is fixed to 0.1 in the fittings of the {\it Chandra} spectra).  By using the {\it xmmpsf} mixing model on the PN spectra of the observation on Dec. 8, 2000, we confirm the difference in column density between regions 10 and 11.  This suggests the hot gas in regions 10 and 11 is likely to be distributed in a more or less symmetric manner with respect to the nucleus, and is not confined to the galactic disk.  Because the near side of the galactic disk is to the SW and the far side to the NE,  we expect strong absorption by gas in the disk toward region 11 and only weak absorption toward region 10, in agreement with our observation. Given the proximity to the active jets, it is possible that some of the hot gas in these two regions is heated by the jets. This hot gas is also one of the key ingredients in the WYC scenario (see their Figs. 5 and 7), in which the gas is ultimately driven into the stellar disk and causes the ``line of damage" that is seen as the anomalous arms. 

Region 12 includes the northern hot spot (end of the currently active radio jet).  The hot spot is not  resolved from the diffuse emission in the {\it XMM-Newton} image.  The total emission from the hot spot is much lower than that of the more extended emission in this region.  The point source $\sim 10\arcsec$ east of the northern hot spot  however,  though unresolved, may slightly alter the {\it XMM-Newton} spectrum from the region.  Region 13 locates on a ``finger"  on the SE arm, at a distance from the nucleus similar to region 12.  The X-ray ``color" suggest the temperature in region 12 is likely lower than that in region 13.  However, the {\it XMM-Newton} spectral parameters are almost identical for the two regions.  This is likely a result of including the hard spectrum point source in region 12.  The low absorption column densities indicate that the emissions from regions 12 and 13  most likely originate from regions in front of the galaxy disk as expected given that the NE side of the galactic disk is the far side.  Region 14 is includes a ``blob" to the east of the nucleus. The {\it Chandra} image shows the spectrum of the west half of the ``blob" is softer than that of the east half.  The ``blob" was reported as a point source (X20) in \markcite{vogler99}{Vogler} \& {Pietsch} (1999). We found the emission is resolved with no indication of a point source in the region.  The low column density  (Table~\ref{tab_aa}) to the region  suggests the ``blob" is located on the near side of  the galactic plane. 

Regions 15--18 are chosen to locate on the X-ray ``filaments"  in the radio ``plateau" regions. Region 16
also includes the far tip of the northern anomalous arm.  The absorption and temperature in region 15 is similar to that of region 1. The low absorption indicates the emission comes mostly from the near side of the galactic disk.   The spectra of regions  16-18, on the other hand, show the highest absorption of all regions, $\sim 5-10$ times the Galactic value, indicating they are on the far side of the galactic disk.  
 
It should be noted that single temperature MEKAL models do not always produce good fits to the spectra, as seen in Table~\ref{tab_aa}.  The fits  for the spectra from regions 1, 5,  13, 15,  16, 17 and 18 are not very good in terms of $\chisq$ per dof.   Most of these regions are in the outskirts of the galaxy or in the  ``plateau".   We introduce a second MEKAL component in these regions, assuming that the absorption and abundance of the component are the same as for the first MEKAL model.  If the second MEKAL is indeed a halo component, the  absorption and abundance of this component may be different from the first MEKAL component.  However, the absorption and abundance of the second MEKAL is poorly constrained with the current data.  Fixing the abundance and absorption does not significantly affect other parameters.  In Table~\ref{tab_2t} we show the best-fit parameters of the double MEKAL model.  By adding a component with temperature $kT \sim 0.2-0.3$~keV, the improvement of $\chisq$ is significant. This is, of course, the $kT \simeq 0.22$~keV gas found in the integrated spectrum of NGC 4258. The normalization of the second MEKAL model is comparable to that of the first component.  The estimated absorption column densities do  not show a significant change from the single MEKAL model. 

These results confirm  the  need for a truly diffuse $kT \sim 0.2-0.3$~keV thermal component. This component is observed mostly in the outer part of the galaxy,  which is consistent with earlier {\it ROSAT} results. The nature of this component, however, is unclear . 

As expected, the {\it Chandra} spectra of the diffuse regions 12---18 agree in general with those obtained using {\it XMM-Newton}, but are noisier due to the low surface brightness.   
 
\subsection{Hot spots
\label{hspot}} 
We examine the X-ray emission from the radio hot spots. The Gaussian smoothed {\it Chandra} soft band contours of the northern and southern radio hot spots overlayed on a {\it Hubble Space Telescope} ({\it HST}) WFPC H$\alpha$+[NII]$\lambda$6583+  image are shown in Fig.~\ref{hst_xray}.  As noticed by \markcite{cecil00}{Cecil} {et~al.} (2000), both radio hot spots are associated with narrow elongated enhancements of H$\alpha$+[NII] emission, suggestive of shock waves, probably bow shocks driven into the ambient gas. The merged {\it Chandra} image provides a clear detection of the northern hot spot (region 20, Fig.~\ref{regions}). The centroid of the X-ray emission lies behind (i.e. to the south of) the H$\alpha$+[NII] emission. The X-ray emission is elongated in the same direction as the optical line emission.  The hot spot is embedded in a plume of diffuse emission in region 12 (Fig.~\ref{regions}).  The background subtracted 0.4--2~keV count rate in a $3.5\arcsec$ radius circular region centered on the hot spot is $(1.1 \pm 0.3)  \times 10^{-3}$~cts/s.  The southern hot spot seen in the radio and H$\alpha$+[NII] images is not detected as a compact X-ray source. However, comparing  the X-ray and the {\it HST} images (Fig.~\ref{hst_xray}),  it is easy to see that the  X-ray isophotes follow the arc-shaped structure seen in H$\alpha$+[NII]. The X-ray emission from the southern bow shock also extends to the north to  form a shell with a radius of $\sim 4.5\arcsec$.  The 0.4--2 keV background subtracted count rate in this circular region is $( 2.0 \pm 0.4)  \times 10^{-3}$~cts/s. We extracted the {\it Chandra} spectra of the hot spots. While noisy, the spectra show little emission above 2 keV in both cases. Because of the small number of counts, spectral modeling is not possible for the hot spots. 

A very soft source roughly on the extension of the southern jet ($77\arcsec$ or 2.7 kpc south of the nucleus, P.A. $176\arcdeg$), previously  named as `SX' in WYC and labeled 19 in Fig.~\ref{regions}, is clearly seen in the {\it Chandra} and {\it XMM-Newton} images, and is speculated to represent another jet-ISM interaction. The source is unresolved in these observations.  The X-ray spectrum using the merged {\it XMM-Newton} observations was extracted from a circular region with a radius of 20\arcsec.  There is little emission above 2~keV. The spectrum can be well described by a bremsstrahlung model absorbed by the Galactic column (allowing the absorption to be a free parameter yields a column density consistent with the Galactic value), with $kT = 0.19_{-0.08}^{+0.07}$~keV, and $f_{0.2-2~keV} = 5.7_{4.0}^{7.0} \times 10^{-15}$~\ergpcmsqps. The 0.2--5~keV luminosity would be $\sim 2.8 \times 10^{37}$~\ergps if the source is in NGC~4258.  

\section{Discussion
\label{discussion}}
\subsection{Geometry of the Anomalous arms}
With spatially resolved {\it XMM-Newton} and {\it Chandra} spectra,  we have shown that the absorption to the NW arm is significantly higher than that to the SE arm.  This difference in absorption has previously been suggested using {\it ROSAT} hardness ratios \markcite{cecil92}({Cecil} {et~al.} 1992). A difference in obscuration has also been invoked to explain the observed difference in the velocity distribution between the SE and NW arms \markcite{rubin90}({Rubin} \& {Graham} 1990). Our result  strongly suggests that the NW arm is partly behind,  while the SE arm is in or in front of  the galactic disk. The simplest interpretation for this is that the arms progressively move out of  the galaxy disk with increasing distance from the nucleus, as suggested in WYC, where the hot gas in the cocoon of the jets hits the galactic disk and shocks the ISM to form the observed anomalous arms.  In this scenario, the shock heated gas in the outer disk is likely to be pushed out of the disk because here the gas is less gravitationally bound than in the inner disk.

The absorbing column to regions 16, 17 and 18 is high, even higher than to the NW anomalous arm. One possible explanation for the extra absorption seen to these regions is that the gas and dust distributions in the disk are non-uniform, as seen in most spiral galaxies. This is directly testable with IR data. The Spitzer IRAC data has recently become available in the archive, which allows us to look at the distribution of warm dust. In Fig.~\ref{x_ir_radio}, we show a color composite of the soft X-ray ({\it Chandra}), 8$\mu$m IRAC, and 1.5 GHz high resolution radio images. It is clear that the warm dust (traced by the 8$\mu$m IR emission) in one of the spiral arms is blocking the soft X-ray emission in the plateau region near the NW arm, while the radio emission is not affected.  This clearly shows that the northern outer arms and the plateau are indeed on the far side of the galactic disk. 

\subsection{Hot gas in the anomalous arms
\label{hotgas}} 
Our X-ray spectra clearly indicate a thermal nature of the X-ray emission from the anomalous arms, and  the gas is very likely to be heated by shocks as suggested in earlier works.  The electron density in the X-ray emitting gas can be estimated using the normalization of the spectra and the volume of the emitting gas. The volumes of spectral extraction regions are calculated 
assuming the regions used for the spectral extractions (Fig.~\ref{regions}) are cylinders with axes along the anomalous arms, except for region 7, for which a spherical shape is assumed.  With the filling factor $f$ as a free parameter, we estimate the electron density $N_e$ (filled circles in Fig.~\ref{ne_vs}) using the PN only normalizations of the single temperature MEKAL models in Table~\ref{tab_aa}.  The density is almost constant ($N_e \sim 0.15f^{-1/2}$~\pcm) along the arms. 
By ignoring the line emissions, we estimate the cooling time of the hot gas, which is 
\begin{equation}
\begin{array}{lll}
t_{cool} & \simeq & \frac{3 N_e kT}{1.4 \times 10^{-27} T^{1/2} N_e^2}  \\
              & \simeq & 1.6 \times 10^8  f^{1/2} \left(\frac{T}{0.6~\rm{keV}}\right)^{1/2} \left( \frac{N_e}{0.15~\rm{cm}^{-3}}\right)^{-1} (\rm{yrs}).
\end{array}
\end{equation}
On the other hand, the evidence of high velocity shock waves found in optical spectra \markcite{cecil00}({Cecil} {et~al.} 2000) allow us to use the best-fit temperature of the hot gas (Table~\ref{tab_aa}) to estimate the velocity of shock waves that heat the gas in the anomalous arms.  Assuming the shocks are strong (e.g. Hollenbach \& McKee 1979)\nocite{hollenbach79},  the estimated shock velocity ranges from 500  to 700~\kmps (Fig.~\ref{ne_vs}).  The time scale for the shock waves to cross the regions is therefore a few times $10^{6}$ yrs, which is much shorter than the cooling time. 

The total volume in our cylindrical regions $V \simeq 1.2 \times 10^{65}$~cm$^{3}$. The total mass of the hot gas is therefore $\sim 1.5 \times 10^7 f^{1/2}$~$\rm{\Msun}$ and the total thermal energy is $3N_ekTV \simeq 4.5 \times 10^{55}f^{1/2}$~ergs.  If the hot gas in the anomalous arms is in thermal equilibrium,  the power needed to heat the gas should be equal to the radiative loss (assumed to be dominated by soft X-ray emission) with $L \simeq 7 \times 10^{39}$~\ergps in the anomalous arms.      
  
\subsection{The nature of the X-ray emission in the hot spots
\label{hsopt}}
To understand the active jets in NGC 4258,  it is of critical importance to understand the radiation mechanism of  the X-ray radiation in the jet hot spots.  Unfortunately, without good quality X-ray spectra, it is hard to elucidate this mechanism.  We  briefly discuss the possibilities. 
 
Synchrotron radiation in the X-ray band has been reported in both radio galaxies and supernova remnants.  In most astrophysical situations, the observed synchrotron spectrum is a power-law ($S \propto \nu^{-\alpha}$).   \markcite{hyman01}{Hyman} {et~al.} (2001) reported a nonthermal radio spectrum of the northern hot spot  with  flux densities of  $0.26 \pm 0.08$ and $0.71 \pm 0.14$~mJy  at 6 and 20 cm,  respectively. The  power-law index  was found to be $0.82 \pm 0.30$.  With the {\it Chandra} count rate, we estimate the radio to X-ray power-law index.  Assuming a single power-law spectrum with $\alpha_x=0.8$ (the X-ray flux is only weakly sensitive to $\alpha_x$),  the {\it Chandra} count rate of the northern hot spot ($\sim 1.1 \times 10^{-3}$ cts/s) translates to an unabsorbed 0.2--2 keV flux  of $\sim 4 \times 10^{-15}$~\ergpcmsqps if the Galactic absorption is assumed, or a flux density of $2.2 \times 10^{-9}$~Jy at 0.4 keV.  Thus the radio to X-ray power-law index is $\sim 0.70$,  which is not inconsistent  with the radio power-law index, suggesting that synchrotron X-ray radiation cannot be ruled out  in the hot spots.  As noted in \S~\ref{hspot},  the X-ray spectra of both radio hot spots seem soft, showing little emission above 2 keV.  Unfortunately, the current X-ray observations do not have sufficient signal-to-noise to  exclude an $\alpha_x \simeq 0.7$ power-law; longer {\it Chandra} observations are needed to address this issue. 

The relativistic electrons in the hot spots can up-scatter  the cosmic microwave background (CMB) photons that dominate the radiation field,  into the X-ray band.  The relative importance of synchrotron and inverse Compton (IC) radiation depends on the relative energy densities of the magnetic field and  that of the CMB photons.  The magnetic field is usually estimated with the minimum energy argument, which, to the order of unity, gives equipartition of energy between magnetic field and the relativistic particles.   If we assume cut-off frequencies of $10^7$  and  $10^{11}$~Hz, adopt the spectral parameters from Hyman et al (2001),  and  choose a radius of the radiation region of the northern hot spot of $2\arcsec$ based on the high resolution radio image from Cecil et al. (2000),  we find an equipartition field $B_m  \approx 21$~$\mu$G  \markcite{ginsburg64}({Ginsburg} \& {Syrovatski} 1964), which is  $\sim 7$ times the equivalent magnetic field of the CMB.  Therefore, it is unlikely that the X-ray radiation is dominated by IC, unless the magnetic field in the hot spots is at least an order of magnitude  lower than $B_m$.  The same argument applies to the southern hot spot.  

Optical spectra of the emission associated with the radio hot spots (Cecil et al. 2000) suggest the gas is heated by shocks with  velocity of $350 \pm 100$ \kmps. Such shocks are capable of heating gas to X-ray emitting temperatures.  This seems to agree with the X-ray spectra if the temperature of the hot gas is below 2 keV. The clear spatial correlation between the X-ray and optical emission (Fig.~\ref{hst_xray}) suggests that the X-ray emission is likely to be associated with these shocks. If the X-ray emission comes from shock heated plasma, and could be modeled as bremsstrahlung absorbed by the Galactic column with a temperature of 0.5~keV (the X-ray colors of the hot spots appear similar to that of the anomalous arms), the count rates would translate to fluxes of  $\sim 3.9  \times 10^{-15}$ and $\sim 7.0  \times 10^{-15}$~\ergpcmsqps, or absorption corrected luminosities of  $2.8 \times 10^{37}$ and  $4.7 \times 10^{37}$~\ergps between 0.4 and 2 keV, for the northern and southern hot spots respectively.  These X-ray luminosities are more than a factor of 100 below that of the anomalous arms (\S 4.2). 

\subsection{Jet energetics and the anomalous arms
\label{jet_energy}}
If the X-ray emission in the hot spots is confirmed to be thermal in nature, the emission is likely to come from the post-shock gas heated by the bow shocks that propagate into the ISM. We favor this interpretation because: 1) The X-ray emission show excellent agreement with the shock structures identified in {\it HST} images; 2) The X-ray color suggests the temperature of the shock heated gas is $\sim 0.5$~keV, which implies a shock velocity of 650~\kmps (e.g. Hollenbach \& Mackee 1979) and which is compatible with the optically measured range of shock velocities ($350 \pm 100$~\kmps, Cecil et al. 2000).  The total power in the shocks can be estimated as follows. Using the X-ray fluxes in the hot spots,  we found the electron density in the post-shock gas to be $0.20f^{-1/2}$ and $0.19f^{-1/2}$~\pcmsq ( $f$ is the filling factor) for the N and S hot spots, respectively (we have adopted spherical volumes with radii of $3.5\arcsec$=122~pc and $4.5\arcsec$=157~pc for the northern and southern hot spots, respectively).  The total power of the bow shocks would be $\frac{1}{2} \rho v_s^3 A$, where  $\rho$ is the density of the ISM, which is 1/4 the density of the shock heated gas in the hot spots in a strong shock, $v_s$ is the velocity of the shock, and $A$ is the area of the hot spot.  Assuming $v_s = 350$~\kmps, the power of the bow shocks is $8.5 \times 10^{38} f^{-1/2}$ and $ 1.7 \times 10^{39} f^{-1/2}$~\ergps for N and S hot spots, respectively.  The total power is therefore $\sim 2.6 \times 10^{39} f^{-1/2}$ ~\ergps.  This power is much higher than the X-ray luminosity of the hot spots ($7.5 \times 10^{37}$~\ergps, see \S~\ref{hsopt}).  In the case of light, non-relativistic jets, the shock heated ISM  mixes with the back flow from the post shock jet gas \markcite{williams91}({Williams} 1991).  The gas then fills a cocoon around the jets which may stabilize the jet flow.  Almost all the power of the jets is  transported  to the cocoon.  It is interesting to note that the power in the bow shocks driven by the jets ($\sim 2.6 \times 10^{39} f^{-1/2}$~\ergps is of the same order as magnitude of the power required to heat the hot gas in the anomalous arms ($\sim 7 \times 10^{39}$~\ergps;  \S~\ref{hotgas}). This is encouraging for the WYC model, in which the anomalous arms are powered by the shock waves driven into the galactic disk by the out-of-plane jets. The shock waves in this scenario could be a result of the cocoon expanding into the ISM; the shock waves transport part of the jet energy in directions transverse to the jet direction.  In particular, the shock waves caused by the expanding cocoon into the disk gas may be responsible for the arc-like features seen in the {\it Chandra} image (regions 4 and 7 in Fig.~\ref{regions}).  These arc-like features are at a similar nuclear distance to the jet hot spots N and SX (regions 20 and 19) and may represent the impact of  ``splatter" from the ends of the jets onto the galactic disk. 

\subsection{Jets and the accretion disk}
If the currently active jets are indeed responsible for heating the anomalous arms, in particular the inner arms in the WYC scenario,  we argue that the power of the jets should be comparable to or greater than the X-ray luminosity of the anomalous arms. This is particularly interesting in NGC~4258 because the accretion power in the AGN can be directly inferred from the disk structure inferred from water maser observations. This allows us to constrain the fractional accretion power which has been injected into the radio jets, which has  important implications for the physics of accretion disk.  

Assuming a thin warped disk illuminated by a central X-ray source \markcite{miyoshi95, greenhill95}({Miyoshi} {et~al.} 1995; {Greenhill} {et~al.} 1995), and using the masing conditions, \markcite{neufeld95}{Neufeld} \& {Maloney} (1995) found an accretion rate $\dot{M} = 7 \times 10^{-5} \alpha$~\Msun \thinspace yr$^{-1}$, where $\alpha \lesssim 1$ is the usual dimensionless viscosity defined in \markcite{shakura73}{Shakura} \& {Sunyaev} (1973). The accretion power is therefore $P = \eta \dot{M} c^2 = 4 \times 10^{41} \alpha$~\ergps, where we have assumed a standard thin accretion disk with efficiency $\eta = 0.1$. On the other hand,  the total X-ray luminosity of the inner anomalous arms (regions 4, 5, 6, 7) is $\sim 4 \times 10^{39}$~\ergps. We further assume that only $\simeq 50\%$ of the energy from the jets has been converted to thermal energy in the disk because only jet power driven down towards the disk can heat the anomalous arms. The power of the jets is thus $\gtrsim 8 \times 10^{39}$~\ergps. This result suggests that the power in the jets is $\gtrsim 2 \alpha^{-1} \%$ of the accretion power.  The power of the jets thus estimated is very similar to the X-ray luminosity of the nucleus.  Note that the argument above also applies to models other than the WYC scenario as long as  the jets from the AGN are the sources of the thermal energy in anomalous arms. The only difference is the factor of 2 introduced above to account for the out-of-plane jets. 

Using a sample of 9 elliptical galaxies, \markcite{allen06}{Allen} {et~al.} (2006) have found recently a tight correlation between accretion and power of the jets (inferred from X-ray emitting gas in the halos/bubbles in the galaxies), 
\begin{equation}
\log (P/10^{43}\rm{\ergps})  = A +B \log( P_{jet}/10^{43}\rm{\ergps})
\label{allen}
\end{equation}
where $P = \eta \dot{M} c^2$,  $\eta = 0.1$,  $A = 0.62 \pm 0.15$ and $B = 0.77 \pm 0.18$. Since the powers of the jets in this sample are typically in the neighborhood of $10^{43}$~\ergps,  it is interesting to test if the relation extends to low luminosity AGNs.  In NGC~4258, we found $\log (P/10^{43}\rm{\ergps}) = -1.4 +\log \alpha$.  On the other hand,  using the power of the jet estimated above, we obtain the RHS $\gtrsim -1.76$. It is obvious that Eqn.~\ref{allen} is satisfied in NGC~4258 to within the uncertainty of the relation.  Should our estimates of the jet power be confirmed in future observations, this result would suggest: 1) While the accretion mode may be very different in elliptical galaxies and low luminosity AGNs, like NGC~4258,  the physics of jet formation is likely to be the same; 2) Since the viscous time scale at 0.13 pc is long ($\sim 10^8$ yr if $\alpha = 0.1$, \markcite{gammie99}{Gammie}, {Narayan}, \&  {Blandford} 1999),  the accretion rate must have been stable over such a long time scale.  
  
\section{Concluding remarks
\label{conclusion}}
In this paper we have examined the spatially resolved archival {\it Chandra} and {\it XMM-Newton} spectra of NGC~4258 in detail. We summarize our main results. 
\begin{enumerate}
\item Using stacked {\it XMM-Newton} spectra, we present a time averaged spectrum of the nucleus of NGC~4258 over 1.5~yrs. Our spectral parameters agree very well with those reported in Hanson06, who use a slightly different method. We confirm the detection of a weak Fe K$\alpha$ line in the combined spectrum, with $E_{line} = 6.39 \pm 0.04$~keV, and $EW = 40 \pm 33$~eV. 
\item The {\it Chandra} spectra of the off-nuclear source J121857.3+471812 show that the source is responsible for the second power-law component detected in the {\it XMM-Newton} spectrum of the nucleus. 
\item The X-ray spectra of the anomalous arms confirm the thermal nature of the X-ray emission. The temperature of the hot gas ($kT \simeq 0.5$~keV) agrees with the idea that the gas is heated by shocks with velocity similar to those implied by optical kinematic measurements \markcite{cecil95}({Cecil} {et~al.} 1995). 
\item The absorption column density to the SE anomalous arm is consistent with the Galactic value, while the column density to the NW arm is significantly higher than the Galactic value. This suggests the SE arm is in the galaxy disk or slightly in front of it,  and, part of the NW arm is on the far side of the galactic disk of NGC~4258. The X-ray emission associated with the NW plateau region appears to be behind the galaxy disk, as inferred from the X-ray column densities and Spitzer IRAC 8$\mu m$ images. 
\item The absorption column density to the diffuse region  SW of the  nucleus is higher than that to the diffuse region  NE of the nucleus, as expected from the fact that the SW side of the galactic disk is the near side. This indicates that a significant part of the hot gas within $\sim 2$~kpc of the nucleus is out of the disk plane.   
\item We confirm that a $ kT \sim 0.2-0.3$~keV diffuse thermal component is present in some regions of the outer part of the galaxy.    
\item We estimate the number density of the X-ray emitting gas along the anomalous arms is $\sim 0.15f^{-1/2}$~\pcm  ($f$ is the filling factor) and the velocity of shock waves that heats the anomalous arms is $600-700$~\kmps. The cooling time of the hot gas is much longer than the dynamical time scale of the shock waves.  
\item The {\it Chandra} image of the radio hot spots shows that the X-ray emission is closely associated with the shock structures identified in {\it HST}  images. While synchrotron radiation cannot be ruled out, it is most likely that the X-ray emission in the hot spots is thermal in nature, and is heated by the shocks responsible for the emission lines seen at optical wavelengths.  
\item Assuming the X-ray emissions from the radio hot spots at the ends of the radio jets come from shock heated gas, we found the power in the shocks is comparable to the power needed to heat the anomalous arms, suggesting that the currently active jets may be responsible for heating the gas in the anomalous arms.  If the AGN jets are indeed the source of thermal energy in the anomalous arms,  we estimate the power of the jets to be $\gtrsim 8 \times 10^{39}$~\ergps.  The power of the jets estimated in this way is $\gtrsim 2 \alpha^{-1} \%$ (where $\alpha$ is the viscosity parameter for the disk) of the accretion power in the disk found by \markcite{neufeld95}{Neufeld} \& {Maloney} (1995).  This seems to agree with  an accretion power versus jet power relation found in elliptical galaxies, even though NGC~4258 has an AGN luminosity three orders of magnitude lower than studied in this relationship.   
\end{enumerate}

We have shown that the currently active jet could be responsible for the formation of the anomalous arms. The X-ray emission from the anomalous arms therefore provides a useful perspective on the energetics of the radio jets and their relation to the accretion disk in NGC~4258.  Our results are consistent with the scenario proposed in WYC, which suggests  the anomalous arms are caused by the impact of shock waves from the jet driven gas in the halo onto the galactic disk.  The arc-like structures in the anomalous arms (Fig.~\ref{true_color}) seen in the {\it Chandra} image could be a natural consequence of hot halo gas shocking the ISM in the galaxy disk.  

To make further progress, good quality X-ray spectra of the radio hot spots are needed. The spectra will be useful to determine the radiation mechanism, and hence gain a better understanding of the jets. Numerical simulations of  highly inclined radio jets in and above a dense galaxy disk  would be desirable to better quantify the energy transport.      

\acknowledgements
We thank Dr. Keith  Arnaud for advice on XSPEC.  This work is supported by NASA through  LTSA grant NAG 513065 to the University of Maryland.  


\vfil\eject\clearpage
\begin{deluxetable}{ccccc}
\tabletypesize{\footnotesize}
\tablewidth{0pt}
\tablecaption{Observations of NGC 4258}
\tablecolumns{7} \tablehead{ 
\colhead{Observatory} &
\colhead{Obs ID} &
\colhead{Detector (filter)\tablenotemark{a}} & 
\colhead{Observation Starting Time} &
\colhead{Exposure time (ks)\tablenotemark{b}}} 
\startdata 
{\it Chandra} & 350 & ACIS-S3 & 2000-04-17 10:52:45 & 14.04 \\  
......        & 1618 & ACIS-S3 & 2001-05-28 21:21:44 & 20.93 \\
{\it XMM-Newton} & 0059140101 & MOS1 (Medium) & 2001-05-06 18:56:45 & 12.00 \\
......           & ......     & MOS2 (Medium) & 2001-05-06 18:56:42 & 12.00 \\
......           & ......     & PN (Medium)   & 2001-05-06 19:36:07 &  7.87 \\
......           & 0059140201 & MOS1 (Medium) & 2001-06-17 16:05:37 & 12.44 \\
......           & ......     & MOS2 (Medium) & 2001-06-17 16:05:37 & 12.44 \\
......           & ......     & PN (Medium)   & 2001-06-17 16:43:12 &  9.05 \\
......           & 0059140401 & MOS1 (Medium) & 2001-12-17 05:26:41 &  6.32 \\
......           & ......     & MOS2 (Medium) & 2001-12-17 05:26:41 &  6.83 \\
......           & ......     & PN (Medium)   & 2001-12-17 06:05:54 &  3.88 \\
......           & 0059140901 & MOS1 (Medium) & 2002-05-22 21:35:06 & 14.86 \\
......           & ......     & MOS2 (Medium) & 2002-05-22 21:35:06 & 14.96 \\
......           & ......     & PN (Medium)   & 2002-05-22 22:08:22 & 10.06 \\
......           & 0110920101 & MOS1 (Medium) & 2000-12-08 22:25:31 & 20.24 \\
......           & ......     & MOS2 (Medium) & 2000-12-08 22:25:31 & 20.26 \\
......           & ......     & PN (Thin1)    & 2000-12-08 23:32:01 & 14.44 \\
\enddata
\tablenotetext{a}{The CCD on which NGC 4258 is observed.}
\tablenotetext{b}{High background intervals have been filtered.}
\label{tab1}
\end{deluxetable}

\begin{deluxetable}{c|l l}
\tablewidth{0pt}
\tablecaption{Model parameters for the X-ray spectra of the nuclear region from {it XMM-Newton} observations}
\tablecolumns{3} 
\tablehead{ 
\colhead{parameter name} &
\colhead{FIT 1} &
\colhead{FIT 2} 
} 
\startdata
$N_{H,gal}$ ($10^{20}$ \pcmsq) &  $2.29_{1.80}^{2.70}$ &
$2.36_{1.48}^{3.36}$ \\
\hline
$kT$ (keV)  & $0.57_{0.55}^{0.59}$ & $0.57_{0.55}^{0.59}$ \\ 
Abundance &   $0.13_{0.10}^{0.16}$ & $0.10_{0.09}^{0.12}$ \\
K\tablenotemark{a} & $3.89_{3.38}^{4.47} \times 10^{-4}$ &
$4.08_{3.67}^{4.49} \times 10^{-4}$ \\
\hline
$N_{H1}$ ($10^{22}$ \pcmsq) & $10.1_{8.48}^{10.3}$ & $8.78_{8.31}^{9.26}$ \\  
$\Gamma_1$ & $1.96_{1.92}^{2.13}$ & $1.66_{1.59}^{1.73}$ \\
$K_1$\tablenotemark{b} & $2.80_{2.49}^{3.16} \times 10^{-3}$ &
$1.82_{1.62}^{2.03} \times 10^{-3}$ \\ 
\hline
$N_{H2}$ ($10^{22}$ \pcmsq) & - & 0.34  (frozen) \\ 
$\Gamma_2$ & $0.10_{-0.20}^{1.94}$ & 1.74 (frozen) \\
$K_2$\tablenotemark{b} & $1.34_{1.04}^{4.93} \times 10^{-5}$ &
$3.36_{2.53}^{3.84} \times 10^{-5}$ \\ 
\hline
$E_{Fe}$ (keV) & $6.37_{6.30}^{6.42}$ & $6.37_{6.31}^{6.42}$ \\ 
$\sigma$ (keV) & $0.05_{0.00}^{0.15}$ & $0.04_{0.00}^{0.14}$ \\
$f_{line}$\tablenotemark{c} & $3.50_{1.86}^{5.53} \times 10^{-6}$ &
$1.82_{1.62}^{2.03} \times 10^{-6}$ \\ 
\hline
$\chi^2/dof$ & 856/836 & 863/837 \\
\enddata
\tablenotetext{a}{$K \equiv 10^{-14} / (4 \pi (D_A (1+z))^2) \int n_e
  n_H dV$, where $D_A$ is the angular size distance to the source
  (cm), $n_e$ is the electron density ($cm^{-3}$), and $n_H$ is the
  hydrogen density ($cm^{-3}$)}
\tablenotetext{b}{defined as the count rate at 1 keV}
\tablenotetext{c}{unit: photons cm$^{-2}$ s$^{-1}$}
\label{tab_nucleus}
\end{deluxetable}
 
\begin{deluxetable}{cccccc|ccccc}
\tabletypesize{\footnotesize}
\tablewidth{0pt}
\rotate
\tablecaption{Model parameters for the X-ray spectra of extended
  regions: Single temperature MEKAL models}
\tablecolumns{11} 
\tablehead{ 
\multicolumn{1}{c}{} &
\multicolumn{5}{c}{ PN } &
\multicolumn{5}{c}{ PN + MOS } \\
\hline
\colhead{Region No.} &
\colhead{$N_{H}$\tablenotemark{a}} &
\colhead{$kT$ (keV)} & 
\colhead{abundance} & 
\colhead{K ($\times 10^{-4}$)\tablenotemark{b}} &
\colhead{$\chi^2/dof$} &
\colhead{$N_{H}$\tablenotemark{a}} &
\colhead{$kT$ (keV)} & 
\colhead{abundance} & 
\colhead{K ($\times 10^{-4}$)\tablenotemark{b}} &
\colhead{$\chi^2/dof$}
}
\startdata 
1  & $0.81_{0.28}^{1.68}$ &$0.41_{0.37}^{0.49}$ &$0.06_{0.04}^{0.08}$
&$3.44_{2.87}^{4.13}$& 97.3/91 & $0.00_{0.00}^{0.53}$ &
$0.53_{0.47}^{0.45}$ & $0.06_{0.05}^{0.07}$ &$2.76_{2.62}^{3.09}$ &
200/163 \\

2  & $1.46_{0.77}^{2.80}$ &$0.42_{0.38}^{0.52}$ &$0.05_{0.04}^{0.07}$
&$2.30_{1.67}^{2.94}$& 56.1/68 & $0.40_{0.00}^{1.05}$ &
$0.49_{0.46}^{0.52}$ & $0.06_{0.05}^{0.08}$ & $1.85_{1.60}^{2.16}$ & 
114/122 \\

3  & $1.11_{0.60}^{1.74}$ &$0.49_{0.47}^{0.51}$ &$0.10_{0.08}^{0.11}$
&$4.59_{4.05}^{5.17}$& 157/176 & $0.31_{0.00}^{0.73}$ &
$0.51_{0.50}^{0.52}$ & $0.11_{0.10}^{0.12}$ & $3.76_{3.46}^{4.10}$ &
287/258 \\

4  & $1.12_{0.51}^{1.76}$ &$0.49_{0.47}^{0.51}$ &$0.12_{0.10}^{0.14}$
&$3.83_{3.36}^{4.33}$& 162/180 & $0.32_{0.00}^{0.77}$ &
$0.51_{0.50}^{0.52}$ & $0.14_{0.12}^{0.15}$& $3.12_{2.88}^{3.45}$ &
275/261 \\

5  & $1.51_{1.02}^{2.05}$ &$0.52_{0.50}^{0.53}$ &$0.13_{0.10}^{0.12}$
&$5.88_{5.35}^{6.48}$& 248/209 & $0.99_{0.66}^{1.40}$ &
$0.53_{0.52}^{0.55}$ & $0.12_{0.11}^{0.13}$ & $5.32_{4.97}^{5.17}$ &
374/301 \\

6\tablenotemark{c}  & $3.05_{2.70}^{3.74}$ &$0.56_{0.54}^{0.57}$ &$0.10_{0.09}^{0.11}$
&$14.2_{13.4}^{15.7}$& 992/986 & $2.85_{2.51}^{3.20}$ &
$0.57_{0.56}^{0.58}$ & $0.10_{0.09}^{0.11}$ & $13.7_{13.0}^{14.3}$ &
1591/1481 \\

7  & $4.69_{2.47}^{6.38}$ &$0.54_{0.50}^{0.58}$ &$0.10_{0.08}^{0.14}$
&$1.74_{1.24}^{2.16}$& 51.8/86 & $4.33_{2.96}^{5.95}$ &
$0.53_{0.50}^{0.56}$ & $0.11_{0.09}^{0.13}$ & $1.66_{1.36}^{2.03}$ &
88.7/139 \\

8  & $3.98_{3.10}^{4.98}$ &$0.50_{0.48}^{0.52}$ &$0.11_{0.09}^{0.13}$
&$5.42_{4.17}^{6.28}$& 194/196 & $3.31_{2.64}^{3.95}$
&$0.52_{0.50}^{0.53}$ & $0.13_{0.12}^{0.14}$ & $4.72_{4.25}^{5.19}$ & 
332/286 \\

9  & $5.65_{3.99}^{9.49}$ &$0.49_{0.40}^{0.51}$ &$0.09_{0.06}^{0.12}$
&$3.24_{2.52}^{5.35}$& 100/113 & $4.36_{3.29}^{5.70}$ &
$0.51_{0.49}^{0.54}$ & $0.12_{0.10}^{0.14}$ & $2.48_{2.11}^{2.98}$ &
205/181 \\

10\tablenotemark{c} & $2.45_{1.99}^{2.97}$ &$0.54_{0.54}^{0.58}$ &$0.09_{0.08}^{0.10}$
&$12.8_{11.8}^{13.8}$& 723/853 & $2.04_{1.73}^{2.37}$ &
$0.57_{0.56}^{0.58}$ & $0.09_{0.08}^{0.10}$ & $1.18_{1.12}^{1.24}$ &
1323/1339 \\

11\tablenotemark{c} & $5.24_{4.44}^{6.25}$ &$0.55_{0.55}^{0.58}$ &$0.09_{0.08}^{0.10}$
&$9.15_{8.22}^{1.04}$& 553/571 & $4.91_{4.30}^{5.54}$ &
$0.58_{0.57}^{0.59}$ & $0.10_{0.09}^{0.11}$ & $8.37_{7.48}^{8.71}$ &
903/928 \\

12 & $1.69_{1.06}^{2.52}$ &$0.51_{0.51}^{0.56}$ &$0.10_{0.09}^{0.12}$
&$3.73_{3.26}^{4.29}$& 126/156 & $1.37_{1.08}^{1.92}$ &
$0.54_{0.52}^{0.56}$ & $0.10_{0.09}^{0.11}$ & $3.61_{3.27}^{3.96}$ &
212/238 \\

13 & $1.59_{1.19}^{2.03}$ &$0.51_{0.51}^{0.54}$ &$0.09_{0.08}^{0.10}$
&$8.54_{7.93}^{9.24}$& 261/224 & $1.05_{0.74}^{1.38}$ &
$0.53_{0.52}^{0.54}$ &$0.10_{0.09}^{0.11}$ & $7.66_{7.24}^{8.10}$ &
461/325 \\

14 & $0.55_{0.07}^{1.41}$ &$0.52_{0.52}^{0.60}$ &$0.11_{0.09}^{0.13}$
&$2.59_{2.29}^{3.16}$& 124/149 & $0.84_{0.00}^{1.34}$ &
$0.54_{0.53}^{0.59}$ & $0.11_{0.10}^{0.12}$ & $2.63_{2.19}^{2.87}$ &
225/226 \\

15 & $0.47_{0.00}^{1.01}$ &$0.46_{0.46}^{0.51}$ &$0.07_{0.06}^{0.08}$
&$4.64_{4.12}^{5.15}$& 202/168 & $0.00_{0.00}^{0.36}$ &
$0.50_{0.48}^{0.51}$ & $0.08_{0.07}^{0.08}$ & $4.19_{4.04}^{4.53}$ &
339/251 \\

16 & $8.21_{7.20}^{10.0}$ &$0.48_{0.48}^{0.54}$ &$0.08_{0.06}^{0.09}$
&$9.91_{8.75}^{12.7}$& 284/208 & $6.21_{5.45}^{7.02}$ &
$0.59_{0.57}^{0.60}$ & $0.10_{0.09}^{0.11}$ & $7.07_{6.42}^{7.78}$ &
466/312 \\

17 & $6.40_{5.12}^{7.69}$ &$0.50_{0.50}^{0.54}$ &$0.10_{0.08}^{0.12}$
&$6.16_{5.12}^{7.67}$& 211/193 & $4.49_{3.74}^{5.21}$ &
$0.58_{0.57}^{0.59}$ & $0.13_{0.12}^{.15}$ & $4.57_{4.11}^{5.74}$ &
345/286 \\

18 & $10.9_{7.74}^{14.2}$ &$0.48_{0.48}^{0.61}$ &$0.06_{0.05}^{0.07}$
&$7.45_{4.64}^{7.61}$& 183/164 & $7.74_{6.72}^{8.92}$ &
$0.60_{0.59}^{0.62}$ & $0.10_{0.09}^{0.12}$ & $4.63_{4.12}^{5.26}$ &
326/259 \\

\enddata
\tablenotetext{a}{in units $10^{20}$\pcmsq}
\tablenotetext{b}{$K \equiv 10^{-14} / (4 \pi (D_A (1+z))^2) \int n_e n_H dV$, where $D_A$ is
          the angular size distance to the source (cm), $n_e$ is the electron
          density ($cm^{-3}$), and $n_H$ is the hydrogen density ($cm^{-3}$)}
\tablenotetext{c}{spectra in 0.2--10~keV band were fitted (see text
  for detail).}
\label{tab_aa}
\end{deluxetable}

\begin{deluxetable}{cccccccc}
\tabletypesize{\footnotesize}
\tablewidth{0pt}
\tablecaption{Model parameters for the X-ray spectra of extended
  regions: Two temperature MEKAL models}
\tablecolumns{8} 
\tablehead{ 
\colhead{Region No.} &
\colhead{$N_{H}$\tablenotemark{a}} &
\colhead{$kT$ (keV)} & 
\colhead{abundance} & 
\colhead{K ($\times 10^{-4}$)\tablenotemark{b}} &
\colhead{$kT$ (keV)} & 
\colhead{K ($\times 10^{-4}$)\tablenotemark{b}} &
\colhead{$\chi^2/dof$} 
}
\startdata 
\multicolumn{8}{c}{ PN } \\
\hline
1  &  $0.35_{0.00}^{1.60}$ & $0.56_{0.49}^{0.62}$ & $0.12_{0.10}^{0.20}$ & 
$1.40_{1.03}^{1.94}$ & $0.19_{0.12}^{0.23}$ & $1.25_{0.94}^{1.65}$ & 77.4/89 \\

5  & $1.29_{0.71}^{1.88}$ & $0.56_{0.53}^{0.59}$ & $0.17_{0.14}^{0.21}$  & 
$3.87_{3.19}^{4.13}$ & $0.21_{0.16}^{0.26}$ & $1.14_{0.89}^{1.65}$ & 221/206 \\

13 & $1.21_{0.87}^{1.77}$ & $0.59_{0.57}^{0.60}$ & $0.15_{0.13}^{0.17}$ &
$4.89_{4.37}^{5.88}$ & $0.23_{0.20}^{0.26}$ & $2.16_{1.76}^{2.72}$ &  202/222 \\

15 & $0.43_{0.00}^{1.32}$ & $0.58_{0.54}^{0.61}$ & $0.19_{0.15}^{0.27}$ & 
$1.87_{1.48}^{2.32}$ & $0.17_{0.14}^{0.20}$ & $1.70_{1.31}^{2.27}$ & 132/166 \\    

16 & $9.48_{7.69}^{11.5}$ & $0.64_{0.60}^{0.67}$ & $0.17_{0.14}^{0.22}$ &
$4.42_{3.50}^{5.48}$ & $0.17_{0.14}^{0.22}$ & $4.20_{3.06}^{6.09}$ & 188/206 \\
 
17 & $6.01_{4.64}^{7.61}$ & $0.61_{0.58}^{0.64}$ & $0.18_{0.14}^{0.23}$ & 
$3.24_{2.54}^{4.08}$ &  $0.18_{0.14}^{0.23}$ & $1.72_{1.15}^{2.90}$ & 174/191 \\
 
18 & $11.4_{8.60}^{14.2}$ & $0.77_{0.62}^{0.86}$ & $0.13_{0.09}^{0.18}$ & 
$2.50_{1.78}^{4.43}$ & $0.30_{0.22}^{0.34}$ & $4.17_{2.66}^{6.20}$ & 144/162 \\
\hline
\multicolumn{8}{c}{ PN+MOS } \\
\hline
1  & $0.00_{0.00}^{0.85} $ & $0.57_{0.55}^{0.60}$ & $0.12_{0.10}^{0.16}$ & $1.33_{1.11}^{1.47}$ & $0.19_{0.14}^{0.22} $ & $1.12_{0.98}^{1.31}$ & 163/161 \\
5  & $0.92_{0.53}^{1.29} $ & $0.57_{0.56}^{0.59}$ & $0.18_{0.16}^{0.20}$ & $3.70_{3.34}^{3.85}$ & $0.20_{0.18}^{0.24} $ & $1.02_{0.83}^{1.23}$ & 331/299 \\
13 & $0.92_{0.58}^{1.27} $ & $0.59_{0.58}^{0.60}$ & $0.16_{0.14}^{0.17}$ & $4.91_{4.47}^{5.39}$ & $0.22_{0.20}^{0.24} $ & $1.82_{1.56}^{2.14}$ & 366/323 \\
15 & $0.00_{0.00}^{0.07} $ & $0.59_{0.58}^{0.61}$ & $0.19_{0.16}^{0.22}$ & $1.86_{1.64}^{2.12}$ & $0.18_{0.15}^{0.20} $ & $1.45_{1.35}^{1.96}$ & 220/249 \\
16 & $9.16_{7.97}^{10.4} $ & $0.63_{0.61}^{0.65}$ & $0.16_{0.14}^{0.18}$ & $4.82_{4.18}^{5.45}$ & $0.22_{0.20}^{0.25} $ & $3.45_{2.77}^{4.42}$ & 334/310 \\ 
17 & $5.81_{4.61}^{6.67} $ & $0.61_{0.59}^{0.64}$ & $0.18_{0.16}^{0.22}$ & $3.31_{2.68}^{3.47}$ & $0.24_{0.20}^{0.26} $ & $1.41_{1.04}^{1.80}$ & 279/284 \\
18 & $10.0_{8.29}^{12.0} $ & $0.77_{0.73}^{0.82}$ & $0.15_{0.12}^{0.19}$ & $2.21_{1.80}^{2.63}$ & $0.32_{0.30}^{0.35} $ & $3.06_{2.29}^{3.49}$ & 261/257 \\

\enddata
\tablenotetext{a}{in units $10^{20}$\pcmsq}
\tablenotetext{b}{$K \equiv 10^{-14} / (4 \pi (D_A (1+z))^2) \int n_e n_H dV$, where $D_A$ is
          the angular size distance to the source (cm), $n_e$ is the electron
          density ($cm^{-3}$), and $n_H$ is the hydrogen density ($cm^{-3}$)}
\label{tab_2t}
\end{deluxetable}

\vfil\eject\clearpage
\enlargethispage*{2000pt}
\begin{figure}
\plotone{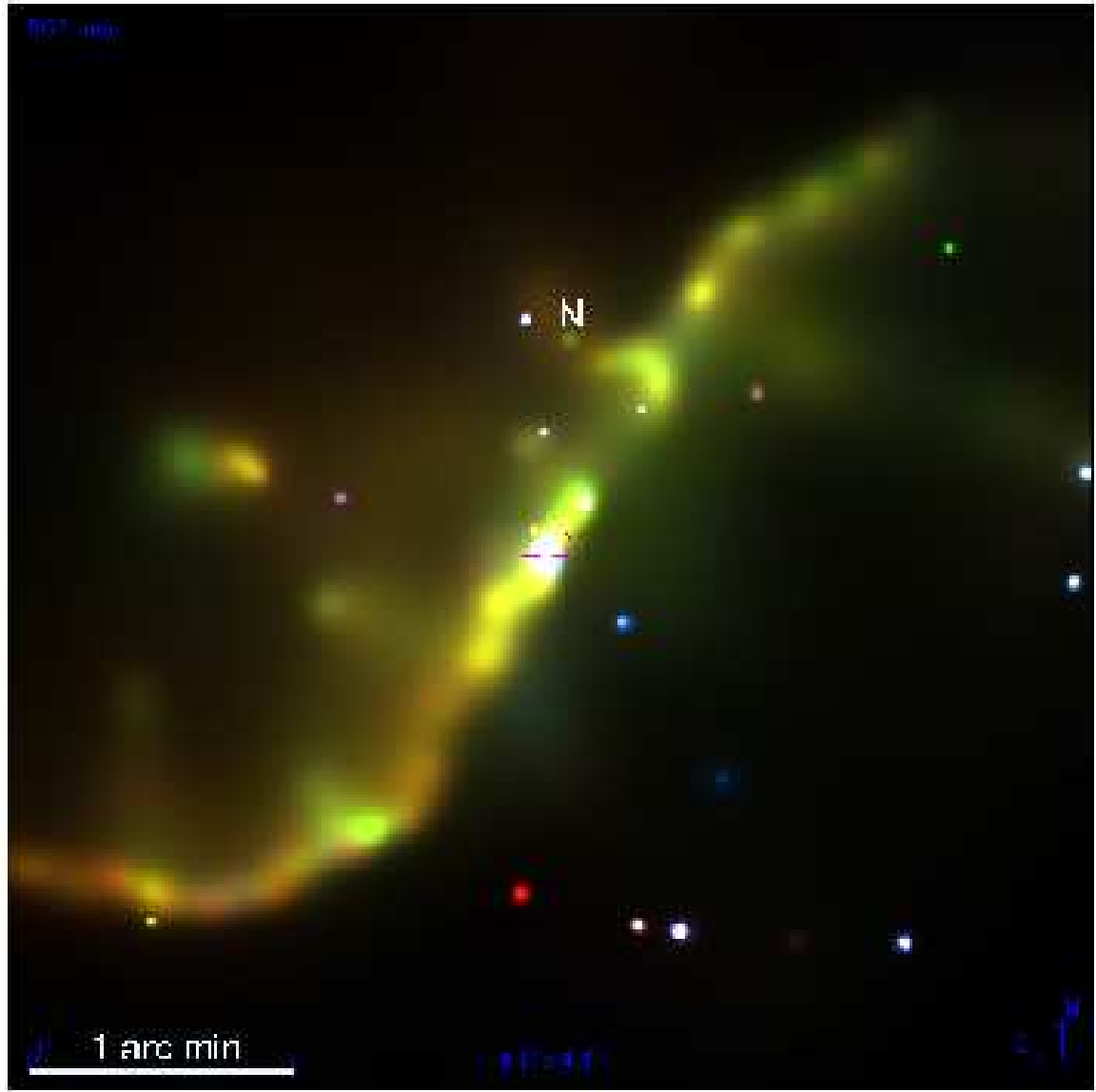}
\caption[true_color.ps] {The adaptively smoothed {\it Chandra} true color image of NGC~4258.  The RGB colors are assigned to represent soft X-ray bands defined as: red: 0.4--0.7~keV; Green: 0.7--1.4~keV; Blue: 1.4--2~keV. The northern hot spot at the end of the radio jet is marked N.   
\label{true_color}}
\end{figure}
\vfil\eject\clearpage
\enlargethispage*{2000pt}
\begin{figure}
\plotone{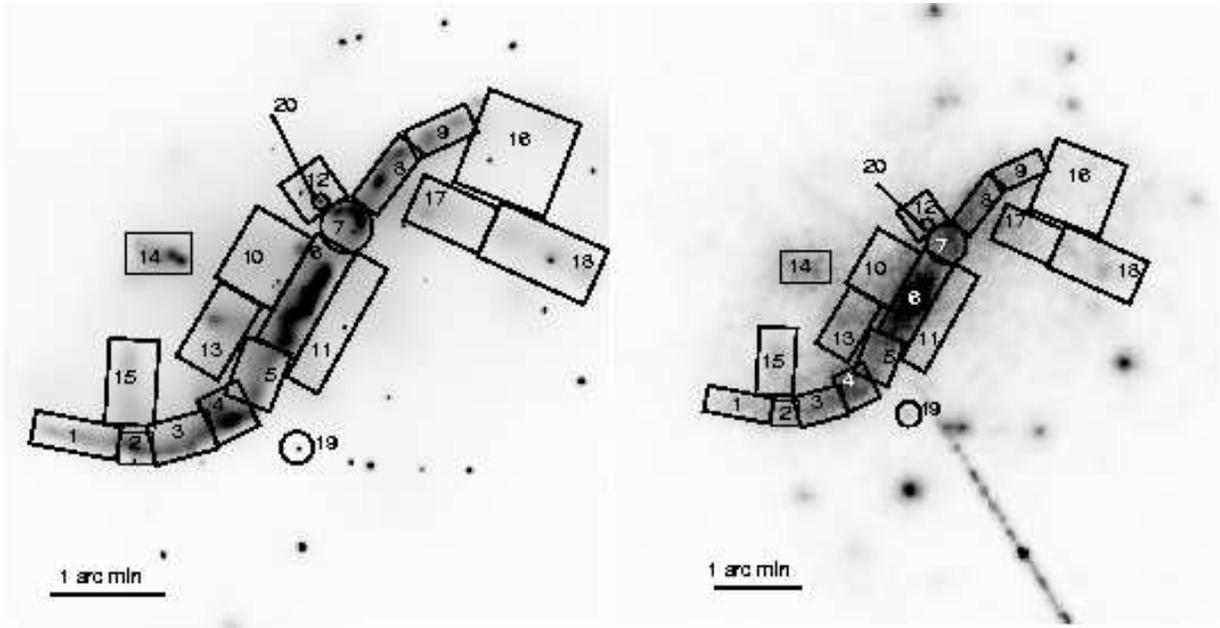}
\caption[XMM_Chandra.ps] {Adaptively smoothed {\it Chandra} 0.4--2 keV (left) and  {\it XMM-Newton} 0.2--2 keV (right) images. The regions for spectral extraction are also displayed.  
\label{regions}}
\end{figure}
\vfil\eject\clearpage
\enlargethispage*{2000pt}
\begin{figure}
\includegraphics[scale=0.6,angle=270]{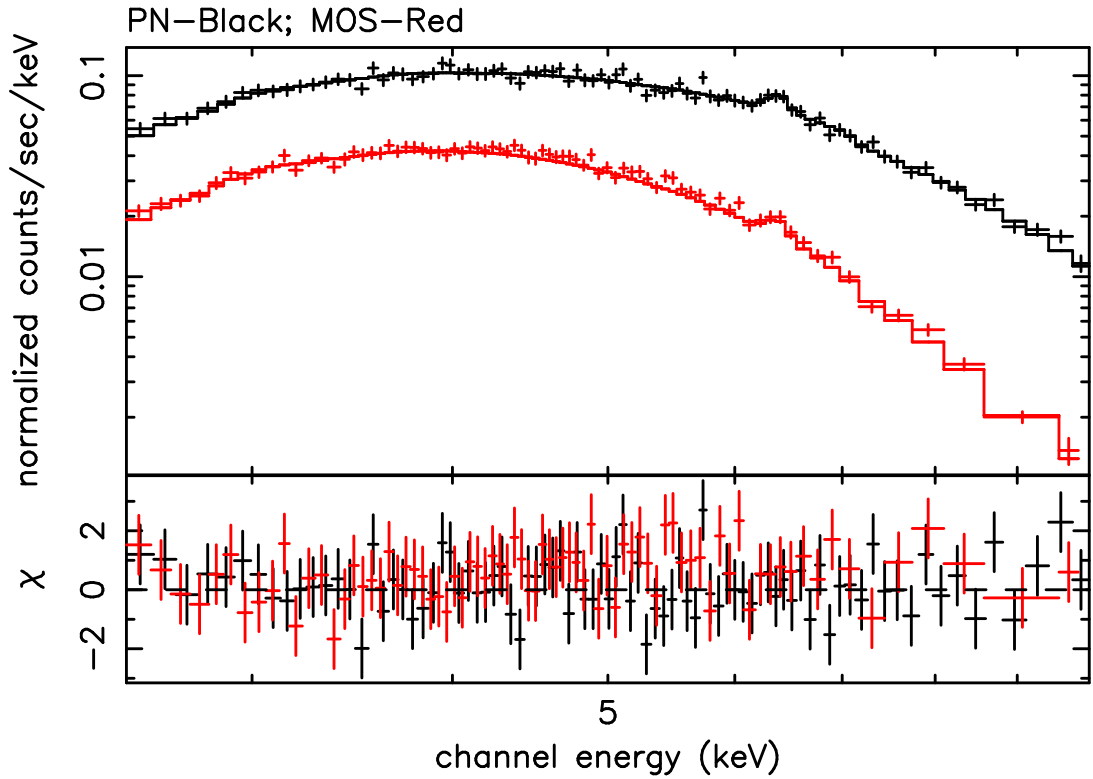}
\caption[nuc_spec]{The stacked 2.2--10~keV PN and MOS spectra and the best-fit model of the nuclear region. The spectra are rebinned so that the signal-to-noise ratio in each bin is $\geq 9$ in XSPEC for better visualization.  An Fe K$\alpha$ emission line is clearly visible in both spectra.
\label{nuc_spec}}
\end{figure}
\vfil\eject\clearpage
\enlargethispage*{2000pt}
\begin{figure}
\includegraphics[scale=0.6,angle=270]{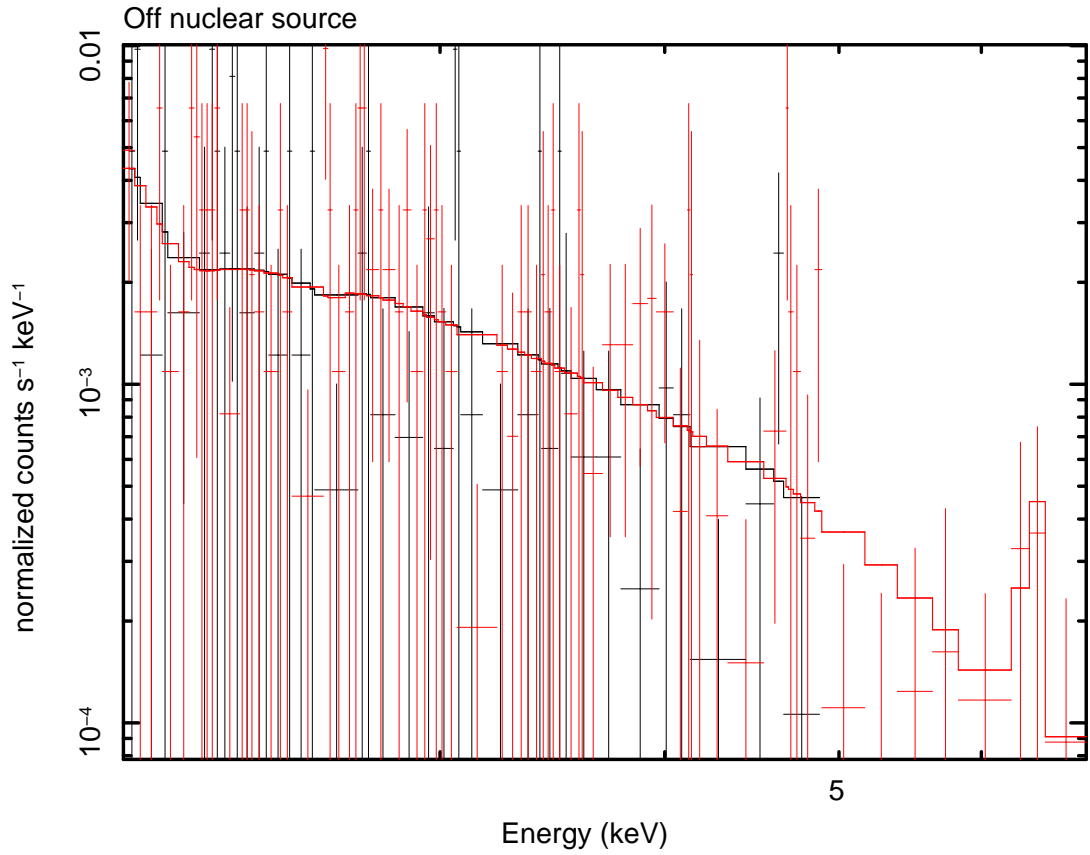}
\caption[fig_offnuc]{The {\it Chandra} spectra of the off-nuclear source. Red: spectrum using the May 28, 2001 observation; Black: spectrum using the April 17, 2000 observation. The spectra are grouped so that each bin has $\geq 3$~cts. A 6.4 keV Fe $K \alpha$ emission line is added to the model.  
\label{fig_offnuc}}
\end{figure}
\vfil\eject\clearpage
\enlargethispage*{2000pt}
\begin{figure}
\includegraphics[scale=0.6,angle=90]{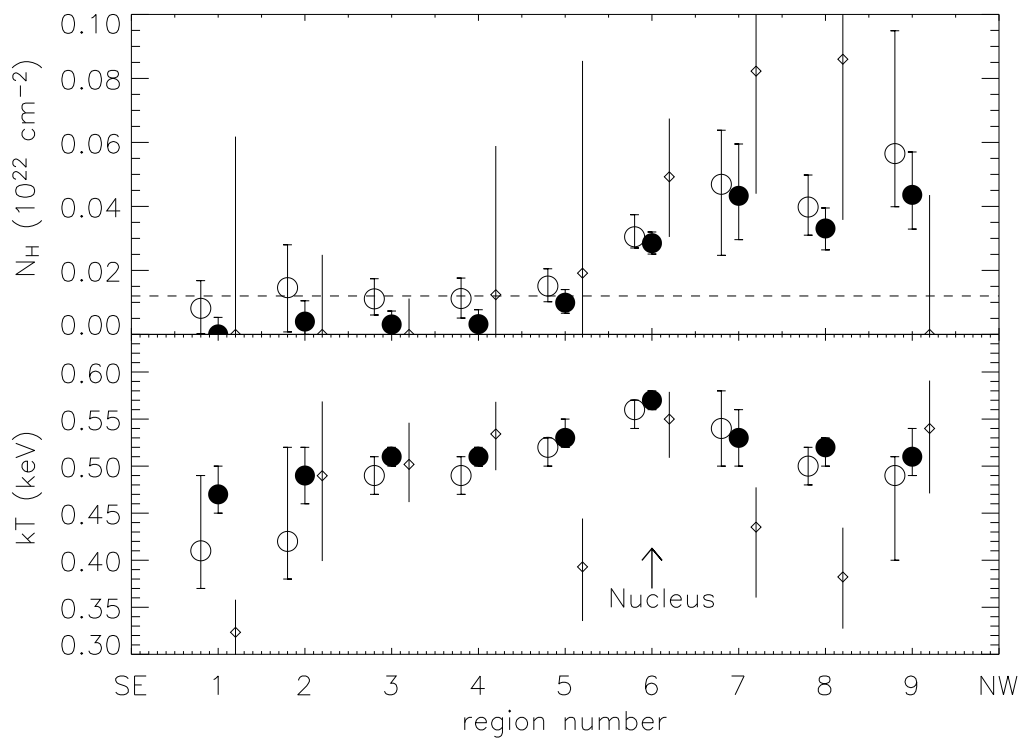}
\caption[aarm_NH_KT]{The absorption column densities and temperatures along the anomalous arms. Region numbers are defined in Fig.~\ref{regions}. The empty and filled circles represent respectively the best-fit using PN spectra only, and both PN and MOS spectra. The small diamonds represent the best-fits using combined {\it Chandra} spectra.  The column density is higher to the NW anomalous arm than to the SE anomalous arm, indicating that the former is on the far side of the galactic disk. 
\label{aarm_NH_KT}}
\end{figure}
\vfil\eject\clearpage
\enlargethispage*{2000pt}
\begin{figure}
\plotone{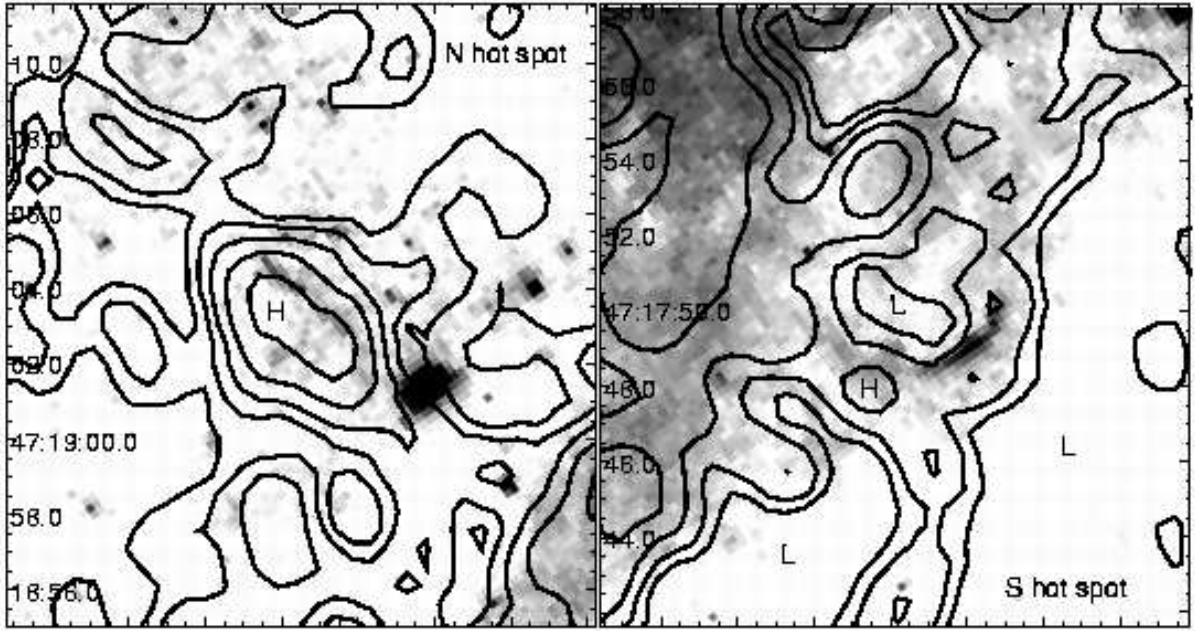}
\caption[hst_xray]{The {\it Chandra} contours of the radio hot spots (left: northern hot spot, marked N in Fig. 1; right: southern hot spot) overlayed on the {\it HST} H$\alpha$+[NII] image \markcite{cecil00}({Cecil} {et~al.} 2000). The southern hot spot is marked S in Fig. 5 of WYC. A clear correspondence between X-ray and optical emissions is seen in these putative shocks.  The high peaks in the X-ray contours  that correspond to the optically identified shocks are  marked ``H",  while the ``valley" behind (i.e. to the north of) the S shock and other areas of low X-ray emission around the S shock are marked with   ``L". 
\label{hst_xray}}
\end{figure}
\vfil\eject\clearpage
\enlargethispage*{2000pt}
\begin{figure}
\plotone{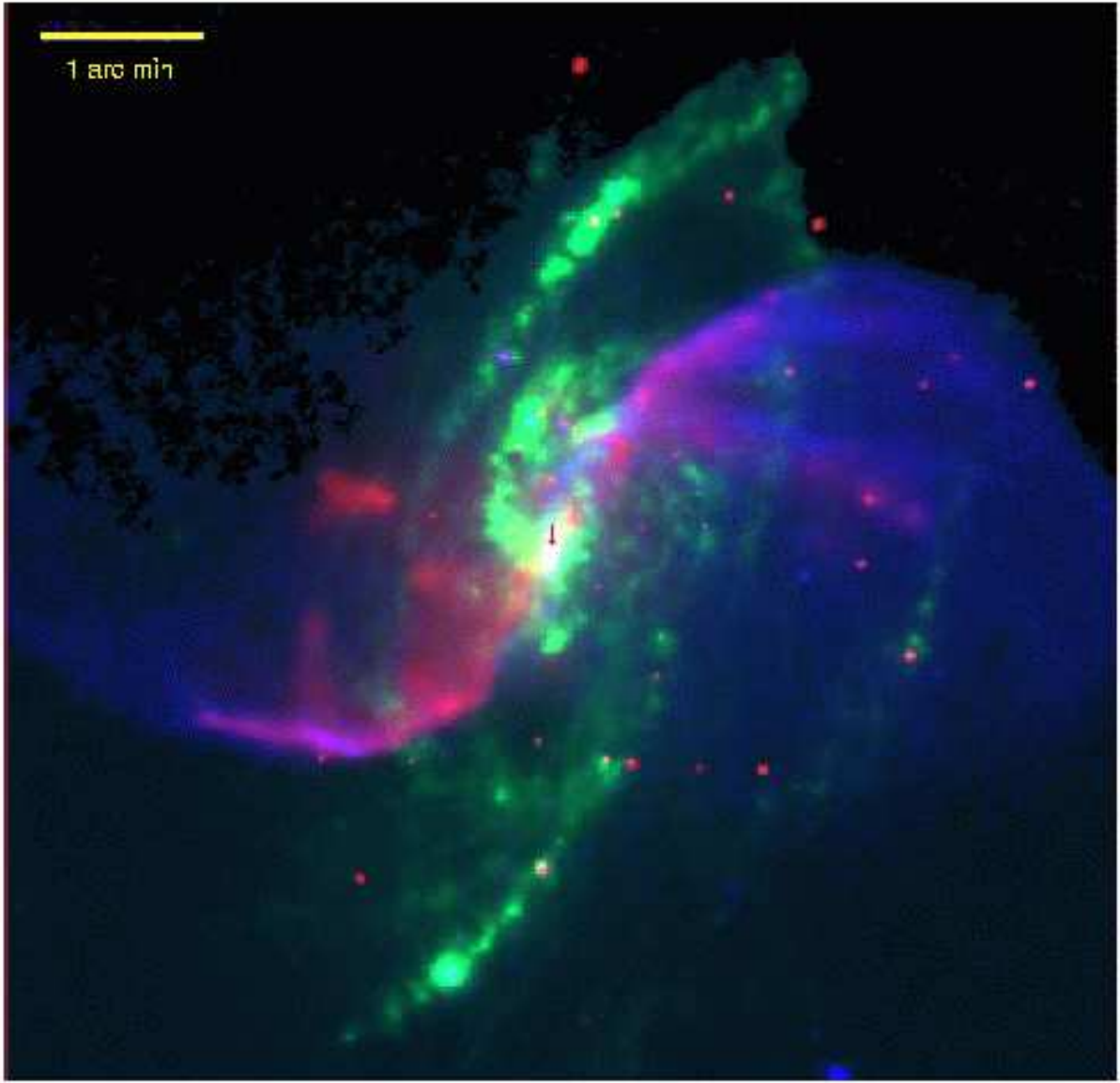}
\caption[x_ir_radio]{The composite  VLA 1.46 GHz \markcite{cecil00}({Cecil} {et~al.} 2000) (blue), {\it Chandra} (red) and {\it Spitzer} 8$\mu$m (green) image of NGC~4258. The X-ray emission at the northern end of the anomalous arms and northern plateau regions is clearly obscured by the warm dust in the galactic disk shown in the 8$\mu$m image. 
\label{x_ir_radio}} 
\end{figure}
\vfil\eject\clearpage
\enlargethispage*{2000pt}
\begin{figure}
\plotone{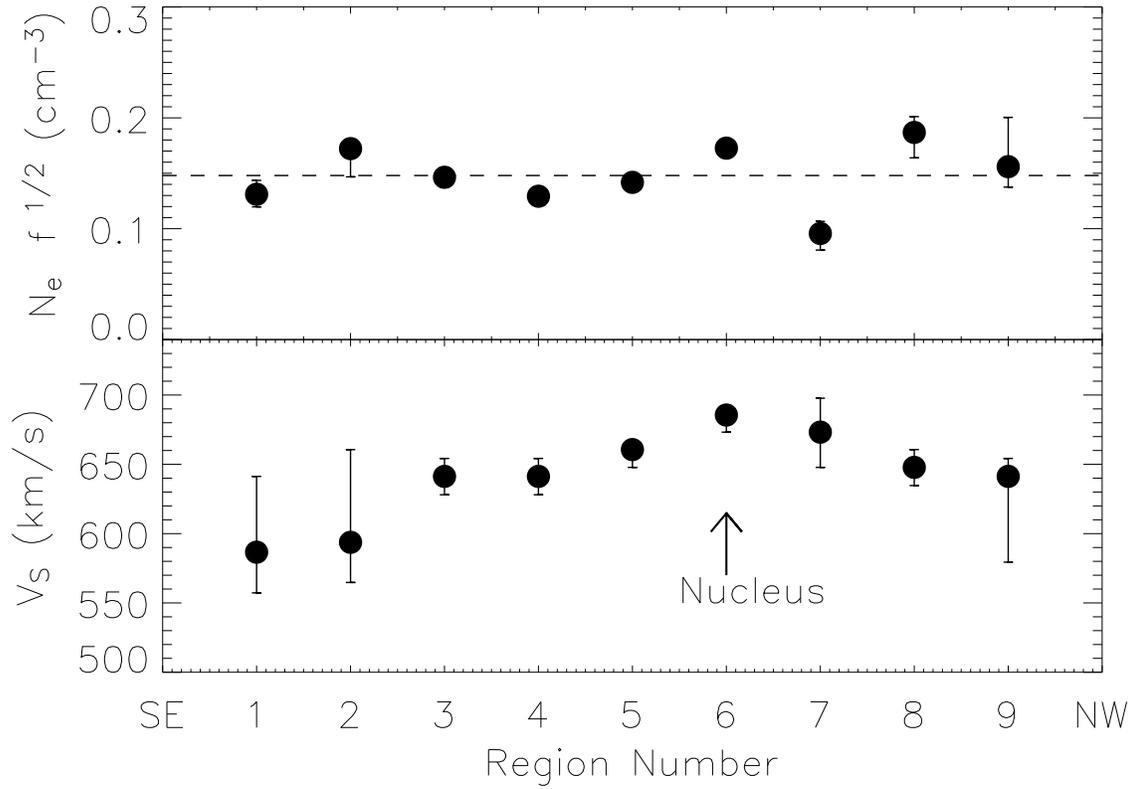}
\caption[ne_vs]{The density of the hot gas in the anomalous arms (upper panel) and the derived shock velocity using the measured X-ray temperatures, assuming the gas is heated by strong shocks (lower panel) . The dashed line is the mean value of $N_e f^{1/2}$.   
\label{ne_vs}}
\end{figure}

\begin{thebibliography}{}

\bibitem[{Allen}, {Dunn}, {Fabian}, {Taylor}, \&  {Reynolds} 2006]{allen06}
{Allen}, S.~W., {Dunn}, R.~J.~H., {Fabian}, A.~C., {Taylor}, G.~B., \&  {Reynolds}, C.~S. 2006, ArXiv Astrophysics e-prints

\bibitem[{Cash} 1979]{cash79}
{Cash}, W. 1979, \apj, 228, 939

\bibitem[{Cecil}, {Greenhill}, {DePree}, {Nagar},  {Wilson}, {Dopita}, {P{\'e}rez-Fournon}, {Argon}, \& {Moran} 2000]{cecil00}
{Cecil}, G., {Greenhill}, L.~J., {DePree}, C.~G., {Nagar}, N., {Wilson}, A.~S.,  {Dopita}, M.~A., {P{\'e}rez-Fournon}, I., {Argon}, A.~L., {et al.} 2000, \apj, 536, 675

\bibitem[{Cecil}, {Morse}, \& {Veilleux} 1995]{cecil95}
{Cecil}, G., {Morse}, J.~A., \& {Veilleux}, S. 1995, \apj, 452, 613

\bibitem[{Cecil}, {Wilson}, \& {Tully} 1992]{cecil92}
{Cecil}, G., {Wilson}, A.~S., \& {Tully}, R.~B. 1992, \apj, 390, 365

\bibitem[{Fiore}, {Pellegrini}, {Matt}, {Antonelli},  {Comastri}, {della Ceca}, {Giallongo}, {Mathur}, {Molendi}, {Siemiginowska},  {Trinchieri}, \& {Wilkes} 2001]{fiore01}
{Fiore}, F., {Pellegrini}, S., {Matt}, G., {Antonelli}, L.~A., {Comastri}, A.,  {della Ceca}, R., {Giallongo}, E., {Mathur}, S., {et al.} 2001, \apj, 556, 150

\bibitem[{Fruscione}, {Greenhill}, {Filippenko},  {Moran}, {Herrnstein}, \& {Galle} 2005]{fruscione05}
{Fruscione}, A., {Greenhill}, L.~J., {Filippenko}, A.~V., {Moran}, J.~M.,  {Herrnstein}, J.~R., \& {Galle}, E. 2005, \apj, 624, 103

\bibitem[{Gammie}, {Narayan}, \&  {Blandford} 1999]{gammie99}
{Gammie}, C.~F., {Narayan}, R., \& {Blandford}, R. 1999, \apj, 516, 177

\bibitem[{Gehrels} 1986]{gehrels86}
{Gehrels}, N. 1986, \apj, 303, 336

\bibitem[{Ginsburg} \& {Syrovatski} 1964]{ginsburg64}
{Ginsburg}, V.~L. \& {Syrovatski}, S.~I. 1964, {The Origin of Cosmic Rays}  (Pergamon Press, New York, 1964)

\bibitem[{Greenhill}, {Jiang}, {Moran}, {Reid},  {Lo}, \& {Claussen} 1995]{greenhill95}
{Greenhill}, L.~J., {Jiang}, D.~R., {Moran}, J.~M., {Reid}, M.~J., {Lo}, K.~Y.,  \& {Claussen}, M.~J. 1995, \apj, 440, 619

\bibitem[{Herrnstein}, {Moran}, {Greenhill},  {Diamond}, {Inoue}, {Nakai}, {Miyoshi}, {Henkel}, \& {Riess} 1999]{herrnstein99}
{Herrnstein}, J.~R., {Moran}, J.~M., {Greenhill}, L.~J., {Diamond}, P.~J.,  {Inoue}, M., {Nakai}, N., {Miyoshi}, M., {Henkel}, C., {et al.} 1999,  \nat, 400, 539

\bibitem[{Herrnstein}, {Moran}, {Greenhill},  {Diamond}, {Miyoshi}, {Nakai}, \& {Inoue} 1997]{herrnstein97}
{Herrnstein}, J.~R., {Moran}, J.~M., {Greenhill}, L.~J., {Diamond}, P.~J.,  {Miyoshi}, M., {Nakai}, N., \& {Inoue}, M. 1997, \apjl, 475, L17+

\bibitem[{Hollenbach} \& {McKee} 1979]{hollenbach79}
{Hollenbach}, D. \& {McKee}, C.~F. 1979, \apjs, 41, 555

\bibitem[{Hummel}, {Krause}, \& {Lesch} 1989]{hummel89}
{Hummel}, E., {Krause}, M., \& {Lesch}, H. 1989, \aap, 211, 266

\bibitem[{Hyman}, {Calle}, {Weiler}, {Lacey}, {Van Dyk},  \& {Sramek} 2001]{hyman01}
{Hyman}, S.~D., {Calle}, D., {Weiler}, K.~W., {Lacey}, C.~K., {Van Dyk}, S.~D.,  \& {Sramek}, R. 2001, \apj, 551, 702

\bibitem[{Icke} 1979]{icke79}
{Icke}, V. 1979, \aap, 74, 42

\bibitem[{Liedahl}, {Osterheld}, \&  {Goldstein} 1995]{liedahl95}
{Liedahl}, D.~A., {Osterheld}, A.~L., \& {Goldstein}, W.~H. 1995, \apjl, 438,  L115

\bibitem[{Makishima}, {Fujimoto}, {Ishisaki}, {Kii},  {Loewenstein}, {Mushotzky}, {Serlemitsos}, {Sonobe}, {Tashiro}, \&  {Yaqoob} 1994]{makishima94}
{Makishima}, K., {Fujimoto}, R., {Ishisaki}, Y., {Kii}, T., {Loewenstein}, M.,  {Mushotzky}, R., {Serlemitsos}, P., {Sonobe}, T., {et al.} 1994, \pasj, 46, L77

\bibitem[{Mewe}, {Gronenschild}, \& {van den  Oord} 1985]{mewe85}
{Mewe}, R., {Gronenschild}, E.~H.~B.~M., \& {van den Oord}, G.~H.~J. 1985,  \aaps, 62, 197

\bibitem[{Mewe}, {Lemen}, \& {van den Oord} 1986]{mewe86}
{Mewe}, R., {Lemen}, J.~R., \& {van den Oord}, G.~H.~J. 1986, \aaps, 65, 511

\bibitem[{Miyoshi}, {Moran}, {Herrnstein},  {Greenhill}, {Nakai}, {Diamond}, \& {Inoue} 1995]{miyoshi95}
{Miyoshi}, M., {Moran}, J., {Herrnstein}, J., {Greenhill}, L., {Nakai}, N.,  {Diamond}, P., \& {Inoue}, M. 1995, \nat, 373, 127

\bibitem[{Neufeld} \& {Maloney} 1995]{neufeld95}
{Neufeld}, D.~A. \& {Maloney}, P.~R. 1995, \apjl, 447, L17+

\bibitem[{Pietsch} \& {Read} 2002]{pietsch02}
{Pietsch}, W. \& {Read}, A.~M. 2002, \aap, 384, 793

\bibitem[{Pietsch}, {Vogler}, {Kahabka}, {Jain}, \&  {Klein} 1994]{pietsch94}
{Pietsch}, W., {Vogler}, A., {Kahabka}, P., {Jain}, A., \& {Klein}, U. 1994,  \aap, 284, 386

\bibitem[{Plante}, {Lo}, {Roy}, {Martin}, \&  {Noreau} 1991]{plante91}
{Plante}, R.~L., {Lo}, K.~Y., {Roy}, J.-R., {Martin}, P., \& {Noreau}, L. 1991,  \apj, 381, 110

\bibitem[{Reynolds}, {Nowak}, \&  {Maloney} 2000]{reynolds00}
{Reynolds}, C.~S., {Nowak}, M.~A., \& {Maloney}, P.~R. 2000, \apj, 540, 143

\bibitem[{Rubin} \& {Graham} 1990]{rubin90}
{Rubin}, V.~C. \& {Graham}, J.~A. 1990, \apjl, 362, L5

\bibitem[{Shakura} \& {Sunyaev} 1973]{shakura73}
{Shakura}, N.~I. \& {Sunyaev}, R.~A. 1973, \aap, 24, 337

\bibitem[{Terashima}, {Iyomoto}, {Ho}, \&  {Ptak} 2002]{terashima02}
{Terashima}, Y., {Iyomoto}, N., {Ho}, L.~C., \& {Ptak}, A.~F. 2002, \apjs, 139,  1

\bibitem[{van Albada} 1977]{vanalbada77}
{van Albada}, G.~D. 1977, \aap, 61, 297

\bibitem[{van Albada} \& {van der Hulst} 1982]{vanalbada82}
{van Albada}, G.~D. \& {van der Hulst}, J.~M. 1982, \aap, 115, 263

\bibitem[{van der Kruit} 1974]{vanderkruit74}
{van der Kruit}, P.~C. 1974, \apj, 192, 1

\bibitem[{van der Kruit}, {Oort}, \&  {Mathewson} 1972]{vanderkruit72}
{van der Kruit}, P.~C., {Oort}, J.~H., \& {Mathewson}, D.~S. 1972, \aap, 21,  169

\bibitem[{Vogler} \& {Pietsch} 1999]{vogler99}
{Vogler}, A. \& {Pietsch}, W. 1999, \aap, 352, 64

\bibitem[{Williams} 1991]{williams91}
{Williams}, A.~G. 1991, {Numerical simulations of radio source structure}  (Beams and Jets in Astrophysics), 342--+

\bibitem[{Wilson}, {Yang}, \& {Cecil} 2001]{wilson01}
{Wilson}, A.~S., {Yang}, Y., \& {Cecil}, G. 2001, \apj, 560, 689

\bibitem[{Young} \& {Wilson} 2004]{young04}
{Young}, A.~J. \& {Wilson}, A.~S. 2004, \apj, 601, 133

\end{thebibliography}
\end{document}